\documentclass[%
aip,%
amsmath,%
amssymb,%
reprint,%
]{revtex4-1}

\usepackage[utf8]{inputenc}
\usepackage{mathptmx}
\usepackage{dcolumn}

\usepackage[T1]{fontenc}
\usepackage{amsmath}
\usepackage{amssymb}
\usepackage{amsfonts}
\usepackage{graphicx}
\usepackage[english]{babel}
\usepackage[raggedright]{sidecap}
\usepackage[caption=false]{subfig}
\sidecaptionvpos{figure}{c}
\usepackage{ragged2e}
\usepackage{bm}
\usepackage{units}
\usepackage{color}
\usepackage[sort&compress]{natbib}
\usepackage{xspace}
\usepackage{tabulary}
\usepackage{enumitem}
\usepackage{multirow}
\usepackage[table,xcdraw]{xcolor}
\usepackage[normalem]{ulem}
\usepackage{hyperref}

\newcommand{\eq}[1]{Eq.~\eqref{#1}\xspace}
\newcommand{\eqs}[2]{Eqs.~\eqref{#1}-\eqref{#2}\xspace}
\newcommand{\eqa}[2]{Eqs.~\eqref{#1} and \eqref{#2}\xspace}

\newcommand{\fig}[1]{Fig.~\ref{#1}\xspace}

\newcommand{\tab}[1]{Tab.~\ref{#1}\xspace}

\newcommand{\quotes}[1]{\lq#1\rq\xspace}

\newcommand{\rnd}[1]{\!\left(#1\right)\xspace}
\newcommand{\sqr}[1]{\!\left[#1\right]\xspace}
\newcommand{\abs}[1]{\!\left|#1\right|\xspace}
\newcommand{\ang}[1]{\!\left\langle#1\right\rangle\xspace}
\newcommand{\cly}[1]{\!\left\lbrace#1\right\rbrace\xspace}
\newcommand{\dif}{\mathrm{d}\xspace}
\newcommand{\kB}{k_{\text{B}}\xspace}
\newcommand{\kT}[1]{\kB T_{#1}\xspace}
\newcommand{\degrees}{{}^{\circ}\xspace}
\newcommand{\Svec}{\mathbf{S}\xspace}
\newcommand{\Qvec}{\mathbf{Q}\xspace}
\newcommand{\rvec}{\mathbf{r}\xspace}
	
\begin{document}
	
	
\title{A preliminary assessment of the sensitivity of uniaxially-driven fusion targets to flux-limited thermal conduction modeling}



\newcommand{\flf}{First Light Fusion Ltd., Unit 9/10 Oxford Industrial Park, Mead Road, Yarnton, Kidlington OX5 1QU, United Kingdom}

\newcommand{\icl}{Centre for Inertial Fusion Studies, Blackett Laboratory, Imperial College, London SW7 2AZ, United Kingdom}

\newcommand{\oxf}{Department of Engineering Science, University of Oxford, Begbroke, Kidlington OX5 1RN, United Kingdom}


\author{D. A. Chapman}
\email{dave.chapman@firstlightfusion.com}
\affiliation{\flf}

\author{J. D. Pecover}
\affiliation{\flf}

\author{N. Chaturvedi}
\affiliation{\flf}
\affiliation{\icl}

\author{N. Niasse}
\affiliation{\flf}

\author{M. P. Read}
\affiliation{\flf}

\author{D. H. Vassilev}
\affiliation{\flf}

\author{J. P. Chittenden}
\affiliation{\icl}

\author{N. Hawker}
\affiliation{\flf}

\author{N. Joiner}
\affiliation{\flf}

\date{\today}


\begin{abstract}
	The role of flux-limited thermal conduction on the fusion performance of the uniaxially-driven targets studied by Derentowicz et al.; Jour.\;Tech.\;Phys.\;18, 465 (1977) and Jour.\;Tech.\;Phys.\;25, 135 (1977), is explored as part of a wider effort to understand and quantify uncertainties in ICF systems sharing similarities with First Light Fusion's projectile-driven concept. We examine the role of uncertainties in plasma microphysics and different choices for the numerical implementation of the conduction operator on simple metrics encapsulating the target performance. The results indicate that choices which affect the description of ionic heat flow between the heated fusion fuel and the gold anvil used to contain it are the most important. The electronic contribution is found to be robustly described by local diffusion. The sensitivities found suggest a prevalent role for quasi-nonlocal ionic transport, especially in the treatment of conduction across material interfaces with strong gradients in temperature and conductivity. We note that none of the simulations produce neutron yields which substantiate those reported by Derentowicz et al.; Jour.\;Tech.\;Phys.\;25, 135 (1977), leaving open future studies aimed at more fully understanding this class of ICF systems.
\end{abstract}

\maketitle


\section{Introduction}
\label{sec:intro}

First Light Fusion (FLF) is investigating a novel approach to controlled inertial confinement fusion (ICF) using efficient, low-cost, projectile-based driver technology. A key characteristic of the concept is that the implosion process initially proceeds with a single, strong shock directed along a single axis, although the dynamics later in time can have convergent, non-planar aspects. As with all routes to ICF, a crucial element of target design and optimization is the validation of the numerical tools, and the multitude of options which underpin them. Sensitivity studies seek to assess the impact of factors such as code configuration choices, numerical methods and the basic properties of materials, and are crucial for enhancing the confidence in simulation-led predictions. Such studies are becoming commonplace in the ICF community \cite{Gaffney_NuclFusion_2013, Gaffney_HEDP_2013, Melvin_PhysPlasmas_2015, Guymer_PhysPlasmas_2015}.

For the planar driver geometry of interest, there are very few experiments accessing the fusion regime against which code benchmarking can be undertaken. One notable class of experiments where the fuel is collapsed directly by a single planar shock are those of Derentowicz et al. \cite{Derentowicz_JourTechPhys_1977, Derentowicz_JourTechPhys_1977b}, which feature a fuel-filled conical cavity in a metal anvil driven by a uniform, planar Mach wave produced by the implosion of a conical liner. Several variations of this experiment have been performed \cite{Krasyuk_QuantElectronics_2005} that consider different driver technologies, such as electrical discharge explosion \cite{Shyam_AtomkernenergieKerntechnik_1984}, direct laser ablation \cite{Vovchenko_JETPLett_1977, Mason_ApplPhysLett_1979} and relativistic electron beams \cite{Bogolyubskii_JETPLett_1976}. The reported output of these targets of $10^{4}-10^{7}$ neutrons is of substantial utility from the perspective of validating both integrated simulations and diagnostic calibration. Of these, the original concept provides the simplest and therefore most valuable validation case.

As for an ICF experiment, the modeling is challenging: compressible hydrodynamics, thermal conduction, viscous drag, radiation generation, transport and absorption and detailed equation of state (EoS) and thermophysical material properties across a wide range of density-temperature space must all be considered. Initial examinations have shown that the most important process for determining fuel energetics is thermal conduction, which acts to homogenize strongly-localised heating resulting from reverberating shock waves.

For plasmas featuring gentle thermal gradients, the rate of conduction loss from a small volume of heated fuel is dictated principally by the thermal conductivities of the electrons and ions; $\kappa_{e}$ and $\kappa_{i}$, respectively. In low-density, high-temperature, systems the electronic contribution dominates due to their higher mobility. In fully-ionized deuterium, ionic conduction contributes equally under nonequilibrium conditions characterized by $T_{i} \gtrsim 4.14\,T_{e}$ \cite{Spitzer_PhysRev_1953, Ferziger_book, Larsen_book}, which may be obtained under the action of strong shock waves. Additionally, electron-ion temperature relaxation directly affects not only the plasma reactivity but also influences the balance between electronic and ionic conduction loss. Although the theoretical descriptions of these properties are well-known under the near-ideal conditions of the fuel, the solid target components are predicted to be driven into warm dense matter (WDM) states, with close-to-solid densities and temperatures of a few $\unit{eV}$. Under such extreme conditions, even basic thermophysical properties can be substantially uncertain \cite{Graziani_book}. 

The system studied in Ref.\;\cite{Derentowicz_JourTechPhys_1977} uses a planar shock in copper to drive the collapse of a deuterium-filled conical cavity with a full internal angle at the tip of $60\degrees$. Between the copper and the fuel is a polythene layer, referred to in this work as the \quotes{coverslip}. The material in which the cavity is made is gold and is referred to here as the \quotes{anvil}. At the point of maximum fuel temperature, which is obtained in the short-lived reverberation produced from interacting shocks reflecting from the tip of the cavity, a transient nonequilibrium state occurs which sets up a thermal wave across the interface between the fuel and interior anvil wall. The resulting heat flow is driven by both electronic and ionic conduction and occurs in the presence of sharp density and temperature gradients between regions with very different conductivities and heat capacities. Accurately capturing the heat flux under such conditions is in general a problem which can only be fully understood using state-of-the-art kinetic codes (see, e.g.\;Refs.\;\cite{Larroche_EuroPhysJourD_2003, Larroche_PhysPlasmas_2012, Taitano_JourComputPhys_2018}) and is one of the many challenges of present concern in the ICF community \cite{Rinderknecht_PlasmaPhysControlFusion_2018}. Reduced kinetic models \cite{Schurtz_PhysPlasmas_2000, Manheimer_PhysPlasmas_2008, Colombant_PhysPlasmas_2008, Cao_PhysPlasmas_2015, DelSorbo_PhysPlasmas_2015, Holec_PhysPlasmas_2018} are available which reproduce many of the features expected from heat flow in the kinetic regime \cite{Hoffman_PhysPlasmas_2015, Brodrick_PhysPlasmas_2017, Sherlock_PhysPlasmas_2017}, although recent experiments \cite{Henchen_PhysRevLett_2018, Henchen_PhysPlasmas_2019} have demonstrated even larger heat flow suppression than predicted. 

Commonly-used simple models, such as local flux-limited thermal conduction \cite{Morse_PhysFluids_1973, Malone_PhysRevLett_1975, Bell_PhysRevLett_1981}, offer a computationally tractable method used in many ICF simulation codes \cite{Meezan_PhysPlasmas_2020}. Although conceptually straightforward, the precise manner in which they are implemented is subject to uncertainty due to the large number of free parameters involved and adequate evidence-based values are seldom available. Of particular importance are: the values of the flux limiter coefficients, the interpolation scheme used to transition between purely diffusive and flux-limited conduction, and the interpolation for evaluating cell-centered quantities, e.g.\;transport coefficients, on cell faces. 

In this work, we report on the initial results of an ongoing simulations-based study investigating the influence of uncertainties in FLF's predictive modeling capability using the experimental results from Ref.\;\cite{Derentowicz_JourTechPhys_1977b} as a benchmark. The scope of this paper is focused on the impact of flux-limited thermal conduction on an idealized model of the target, as this is currently believed to have the largest effect on the fusion output. With suitably well-converged simulations, we establish a reference case against which the impact of configuration changes can be quantifiably measured using a few simple metrics. The basic form of the conduction operator is described, with emphasis given to the main sources of uncertainty arising from free choices in its implementation. 

We perform a sensitivity study to the microphysics relevant to conduction for the main components of the simulation by applying scaling factors, both uniformly and through a more realistic \quotes{targeted} approach based on the WDM parameter introduced by Murillo \cite{Murillo_PhysRevE_2010}. Whilst the performance of the target can be significantly affected using large, uniformly-applied scaling factors, only a negligible sensitivity is found if the targeted approach is used. This is because the state of fuel is always close to ideal despite being eventually compressed by a factor of almost $2000$. We further assess the impact of the free parameters in the conduction operator setup, with the results showing that the choice of how the heat fluxes are evaluated of cell faces in the discretized system has a crucial influence on both the target evolution and fusion performance. More generally, we find that any change resulting in strong restriction of the ionic heat flux leads to substantial changes in the operation of the target, suggesting a prominent role for nonlocal ion transport. Conversely, we find the electronic heat flow to be well-described by local diffusion. Given the low uncertainty in the thermophysical properties of the fuel under the conditions produced, we conclude that an improved understanding of ionic heat transport in the kinetic regime, especially for heterogeneous systems such as material interfaces \cite{Stanton_PhysRevX_2018}, will be crucial if we are to use this experimental platform as a means to validate simulations of relevance to FLF's mission. 


\section{Simulation setup}
\label{sec:sim_setup}

\subsection{Idealized target model}
\label{subsec:idealised_target_model}

\begin{figure*}
	\subfloat{
		\includegraphics[width=0.82\textwidth]
		{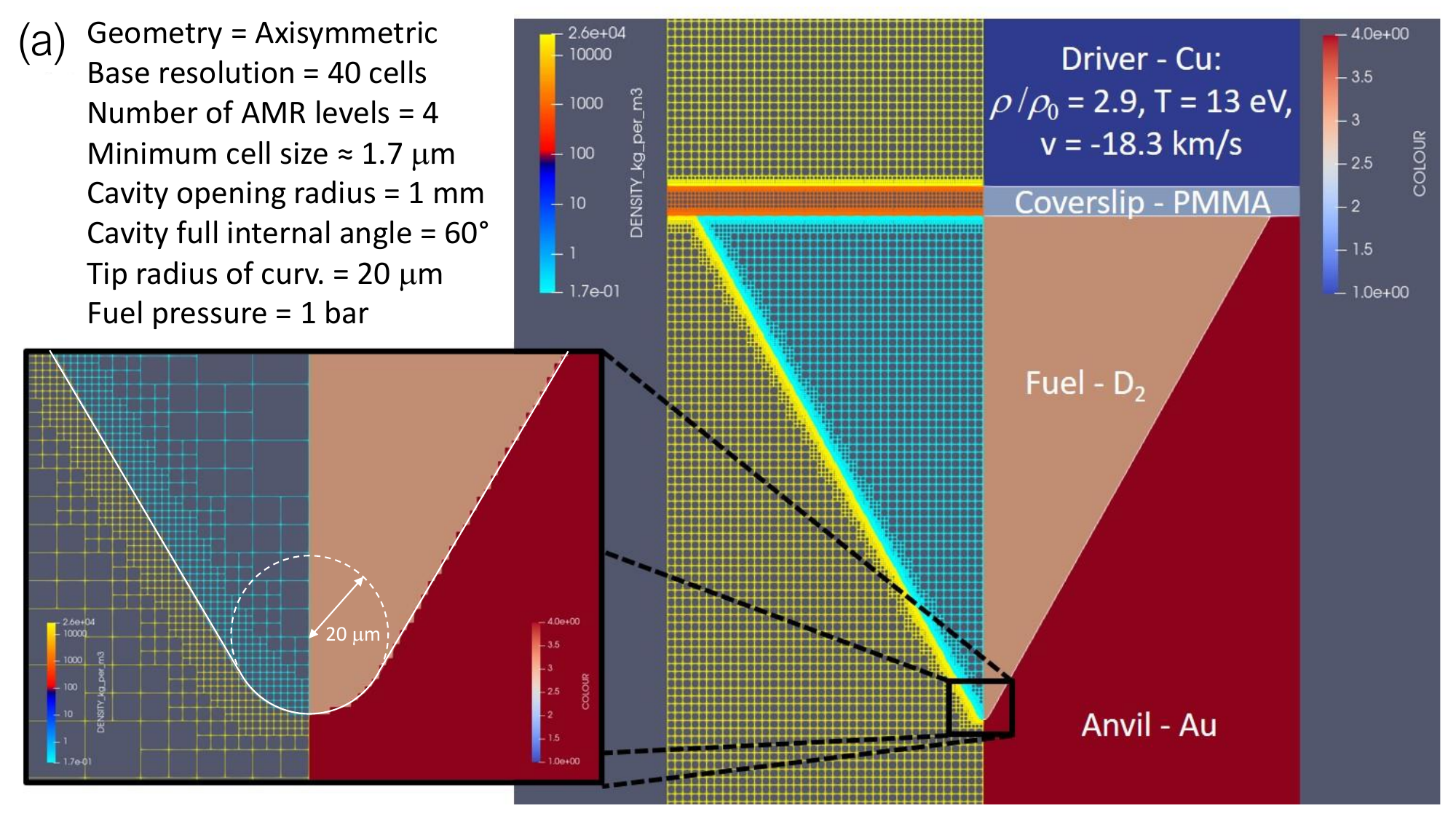}
		\label{fig:ideal_model_setup}
	}
	\qquad
	\subfloat{
		\includegraphics[width=0.57\textwidth]
		{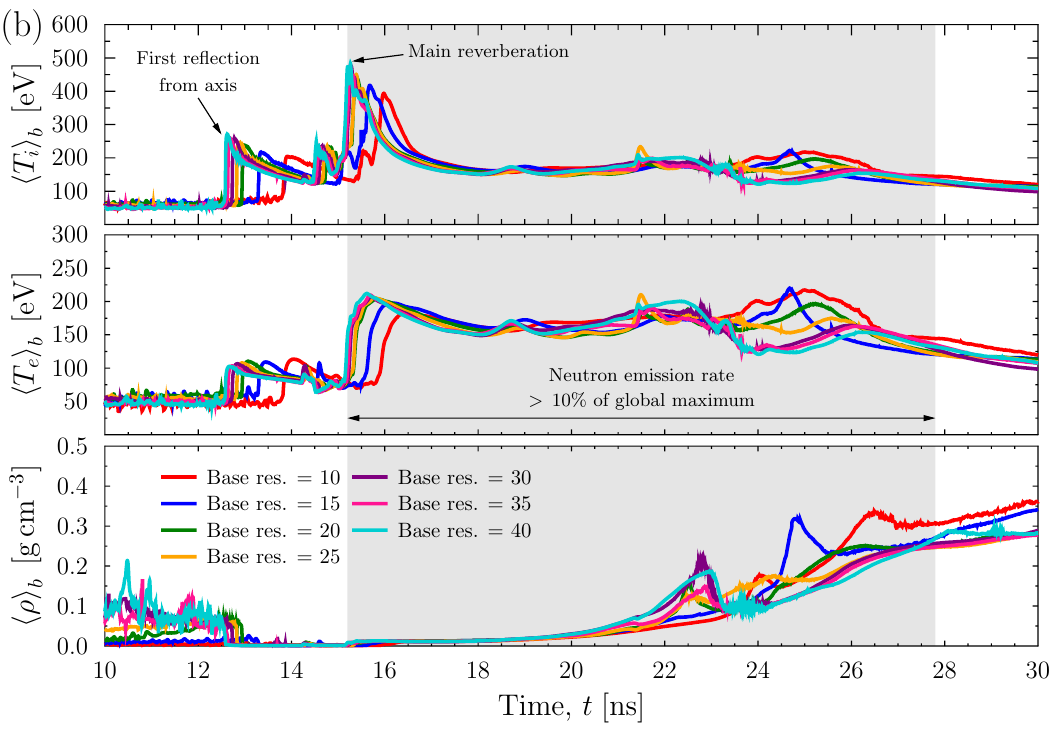}
		\label{fig:bwa_variable_profiles}
	}
	\subfloat{
		\includegraphics[width=0.3\textwidth]
		{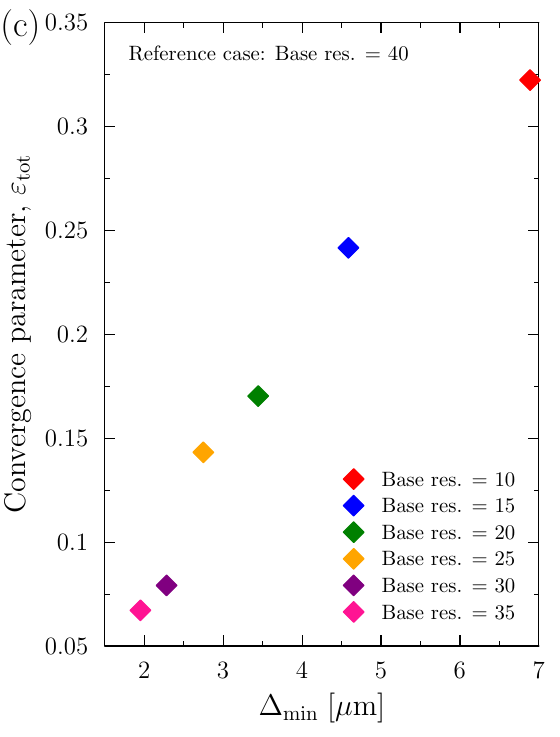}
		\label{fig:convergence_parameter}
	}
	\caption{
		(a): Simulation geometry and EGrid zoning for the idealized model, featuring a shocked copper driver, PMMA coverslip, deuterium fuel and gold anvil. The left-hand part of the plot shows a logarithmic map of mass density whilst the right-hand side shows the material \quotes{color} indicator. The tip of the conical cavity is rounded with a radius of curvature of $20\;\unit{\mu m}$, as shown in detail in the inset. (b): Time histories of the BWA ion (upper panel) and electron (middle panel) temperatures and mass density (lower panel). The regions with the gray background indicate the time window over which simulation convergence is assessed. (c): Convergence metric $\epsilon_{\text{tot}}$ as a function of minimum cell size.
	}
\end{figure*}

The present work is undertaken using Hytrac; one of the two in-house radiation-hydrodynamics codes developed by FLF. The code is based on the front tracking approach \cite{Jian_JComputPhys_2006} and also implements cell-based adaptive mesh refinement, similar to, e.g., RAGE \cite{Gittings_ComputSciDisc_2008}. The multi-physics model includes two-temperature thermal conduction, viscous momentum and energy fluxes \cite{Vold_PhysPlasmas_2017} and radiation transport via the $P_{1/3}$-AFL method \cite{Mihalas_JQSRT_1982, McGlinchey_thesis_2017}. The code further supports either analytic or tabulated models for material EoS, plasma microphysics and non-thermal fusion reactivity.

The work presented in Refs.\;\cite{Derentowicz_JourTechPhys_1977, Derentowicz_JourTechPhys_1977b} shows that the drive into the fuel is well-approximated with a single, uniform shock into the plastic coverslip. The pressure in the copper is inferred from shock velocity measurements to be $46\;\unit{Mbar}$ ($\rho \approx 26\;\unit{g\,cm^{-3}}$ and $T \approx 13\;\unit{eV}$), which leads to a shock in the coverslip with a pressure of $13\;\unit{Mbar}$ ($\rho \approx 3.8\;\unit{g\,cm^{-3}}$ and $T \approx 7.4\;\unit{eV}$). The latter is modeled as PMMA as a surrogate for polythene due to difficulties matching the principal Hugoniot using the FEOS EoS model \cite{Faik_ComputPhysCommun_2018}. Since the densities of the two materials are sufficiently similar, this compromise is believed to be reasonable.

The target mesh geometry used for our simulations is shown in \fig{fig:ideal_model_setup}. The number of AMR refinement levels is 4, giving a minimum cell size of $1.7\;\unit{\mu m}$ on the Eulerian grid (EGrid) using a base resolution of 40 cells. The inset shows how the cavity tip has had a radius of curvature of $20\;\unit{\mu m}$ added to account for expected manufacturing tolerances. The shocked copper region, denoted as the \quotes{driver}, at the top of the domain is initialised from the thermodynamic state resulting from shocking initially ambient copper to the experimentally inferred pressure along the principal Hugoniot. All other materials are initialised at a pressure of $1\;\unit{bar}$ and their appropriate ambient densities to produce initially static interfaces.

In these simulations, the viscous flux and radiation transport operators are not used. In the case of viscous effects, this is because the predominant influence is on the damping of the incident shock and not the later stages of the cavity collapse, from which the bulk of the yield is expected. Full radiation-hydrodynamics modelling suggests the internal dynamics may be influenced by bremsstrahlung emission from the fuel to a small degree through radiative ablation of the internal cavity wall. However, it is believed that thermal conduction loss is vastly more important to the fuel energetics. Disabling these operators also makes tractable a larger study focused on the physics of conduction, which is our principal focus here. More details of the cavity collapse dynamics and the impact of the full range of available multi-physics options will be discussed in future work.

\subsection{Assessment of convergence}
\label{subsec:sim_convergence}

\newcommand{\bwa}[2]{\ang{#1}_{b}\rnd{#2}\xspace}	

In order to assess the impact of changes in the simulation configuration on the performance of the study, it is crucial to ensure that the simulations are well-converged in both time and space. We pursue convergence through variation of the EGrid base resolution at a constant refinement level and comparing the difference between key simulation probes for progressively higher-resolution runs. 

The probes used for this purpose track the burn-weighted average (abbreviated in this work as BWA) histories of the ion and electron temperatures and the mass density
\begin{align}
\label{eq:bwa_variable_def}
\bwa{X}{t}
= &\,
\dfrac{
	\displaystyle\int_{V}\dif\rvec\,\,	\dfrac{\partial^{2} Y_{\text{n}}}{\partial V \partial t}\,
	X(\rvec,t)
}{
	\displaystyle\int_{V}\dif\rvec\,\,
	\dfrac{\partial^{2} Y_{\text{n}}}{\partial V \partial t}
}
\,.
\end{align}
Here, $X = \cly{T_{i}, T_{e}, \rho}$ stands for the quantity of interest, $\partial^{2} Y_{\text{n}}/\partial V\partial t$ represents the total neutron yield emission per unit volume per unit time and $V$ denotes the volume occupied by the fusion fuel. The convergence parameter $\epsilon$ for variable $X$ calculated from a simulation with minimum EGrid cell size $\Delta_{\min}$ is then defined relative to a suitable reference case as
\begin{align}
\label{eq:convergence_def}
\epsilon_{X}
= &\,
\sqr{
	\frac{\sum_{i_{\min}}^{i_{\max}}
		\rnd{\bwa{X}{t_{i}; \Delta_{\min}} - \bwa{X}{t_{i}; \Delta_{\min}^{\text{ref}}}}^{2}}{i_{\max}-i_{\min}}
}^{1/2}
\,.
\end{align}
Here, the index $i$ runs between times bounded by the onset ($t_{i_{\min}}$) and secession ($t_{i_{\max}}$) of neutron production. These are respectively defined as the earliest and latest times where the neutron production rate exceeds $10\%$ of its peak value. The convergence metric is then defined as the quadratic average over all variables, i.e.\;$\epsilon_{\text{tot}} = \rnd{\sum_{X}\epsilon_{X}^{2}/\sum_{X}}^{1/2}$. For pure deuterium fuel at the modest temperatures predicted, it is reasonable to ignore the effect of fuel depletion and secondary reactions involving fusion products. Thus, we may approximately write \cite{Atzeni_book}
\begin{align}
\label{eq:volumetric_neutron_emission_rate}
\frac{\partial^{2} Y_{\text{n}}}{\partial V \partial t}
\approx &\,
\frac{n_{\text{D}}^{2}}{2}
\ang{\sigma v}_{d(D,{}^{3}\text{He})\text{n}}
\,.
\end{align}
In \eq{eq:volumetric_neutron_emission_rate} the thermal reactivity $\ang{\sigma v}$ is given by fits due to Bosch and Hale \cite{Bosch_NuclFusion_1992}. 

A strong convergence trend can be seen in both the timing and shape of the peaks in $\ang{T_{i}}_{b}$ and $\ang{T_{e}}_{b}$ with increasing base resolution (\fig{fig:bwa_variable_profiles}), indicating that the hydrodynamics of the cavity implosion are properly captured in our simulations. This is further exemplified in \fig{fig:convergence_parameter}. The two main spikes labeled in the BWA ion temperature profile correspond to the first reflection of the incident shock from the simulation axis and the short-lived two-temperature state produced by the reverberation created from multiple interacting shock reflections from the tip of the cavity; this event is hereafter referred to as the \quotes{main reverberation}. The subsequent weaker reverberations in the collapsing cavity produce a long-lived plasma in which conduction losses are roughly balanced by the heating due to compression. The BWA electron temperature profile does not show the same strongly peaked structure since the electrons are heated isentropically through shocks, as per the scheme used in the FLASH code \cite{Fryxell_AstrophysJSupplSeries_2000}. Nonequilibrium states with $T_{i}/T_{e} > 3$ are achieved in this period before equilibration to a temperature of $T \sim 160-180\;\unit{eV}$, which persists for around $10\;\unit{ns}$ as the fuel continues to be axially compressed. The role of ion conduction in dissipating the energy imparted to the fuel at the tip of the cavity is therefore likely to be important to the overall heat loss, as is the rate of electron-ion temperature equilibration. The strength of the observed trends indicates that our results at a base resolution of 40 cells (smallest cell size of $1.7\;\unit{\mu m}$) provides a suitable reference case for quantifying the impact of changes due to conduction physics, without conflation with issues related to lack of convergence.


\section{Conduction modeling in ICF simulations}
\label{sec:conduction_operator}

\subsection{Flux-limited thermal conduction}

It is well-known that the modeling of losses due to thermal conduction constitute a crucial aspect of the power balance in prospective ICF systems \cite{Atzeni_book}. A commonly-used implementation strategy for the conduction operator in radiation-hydrodynamics codes follows the local flux-limited diffusion approach to Fourier's law
\begin{align}
\label{eq:fourier_heat_flux}
\Qvec_{\text{Four}} 
= &\,
-\kappa \nabla T
\,,
\end{align}
the magnitude of which is restricted to a fraction, $\alpha$, of the free-streaming limit \cite{Malone_PhysRevLett_1975}
\begin{align}
\label{eq:max_heat_flux}
Q_{\max} 
= 
\alpha n \, v_{\text{th}} \, \kT{}, 
~~~ 
v_{\text{th}} 
= 
\sqrt{\kT{}/m}
\,,
\end{align}
in order to prevent unphysically large fluxes from arising in regions with steep temperature gradients. Such conditions arise routinely near the ablation front of laser-irradiated solids, e.g.\;in short-pulse laser-driven targets \cite{Brown_PhysRevLett_2011,Hoarty_PhysRevLett_2013} or the heating of hohlraum walls at the NIF \cite{Meezan_PhysPlasmas_2020}. In the present study, sharp temperature gradients are expected to occur in proximity to strong shocks and also at material interfaces; especially between the shock-heated fuel and the confining anvil.

The flux-limited model represents a purely phenomenological correction to purely diffusive energy transport. Whilst it can approximately capture the inhibition of the local flux through tuning of the value of $\alpha$, it fails to account for the preheating effect arising from long-range (near collisionless) propagation of the high-energy tail of the particle distribution. The impact of nonlocal transport is now of growing concern in the ICF community \cite{Hoffman_PhysPlasmas_2015, Rinderknecht_PlasmaPhysControlFusion_2018} and has driven the development of numerous high-fidelity kinetic simulation tools \cite{Larroche_EuroPhysJourD_2003, Huang_PhysPlasmas_2017} and accurate reduced models, such as the popular SNB approach \cite{Schurtz_PhysPlasmas_2000}. Unfortunately, these more accurate capabilities are seldom practical for implementation in fully-integrated simulations and the flux-limited approach remains in common usage. Even for this simple model, however, there are a number of free parameters to set and numerical choices to be made; independently for the electrons and ions.

\subsection{Flux limiter coefficients}

For the electrons, the value of $\alpha_{e}$ is typically tuned to match experimental data or designed to match results from high-fidelity simulations, such as kinetic codes, with $\alpha_{e} \approx 0.05$ being a common choice informed by early Vlasov-Fokker-Planck (VFP) results \cite{Bell_PhysRevLett_1981}. Despite being informed by numerical results, a broad range of values may be needed to match different experiments, with values up to $\alpha_{e} = 0.1-0.15$ being required (see, e.g.,\;Refs.\;\cite{Goldsack_PhysFluids_1982, Rosen_HEDP_2011}). On the other hand, recent work by Meezan et al. \cite{Meezan_PhysPlasmas_2020} has shown that a flux-limited approach with a lower value of $\alpha_{e} = 0.03$ can qualitatively explain results from several independent diagnostics simultaneously. Thus, a wide range of values for this parameter can be substantiated for sensitivity studies.

\begin{figure*}
	\subfloat{\raisebox{0.05cm}{%
		\includegraphics[width=0.38\textwidth]
		{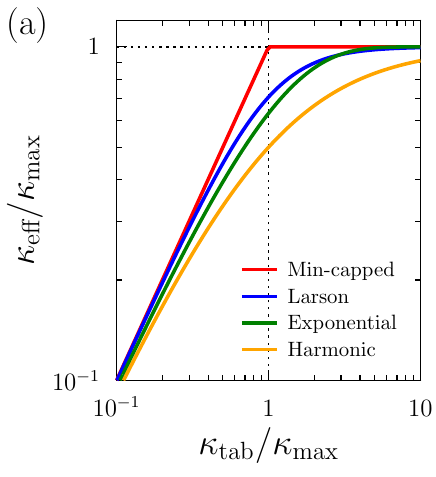}
		\label{fig:flux_limiter_interpolation_models}%
	}}
	\subfloat{\raisebox{0cm}{%
		\includegraphics[width=0.55\textwidth]
		{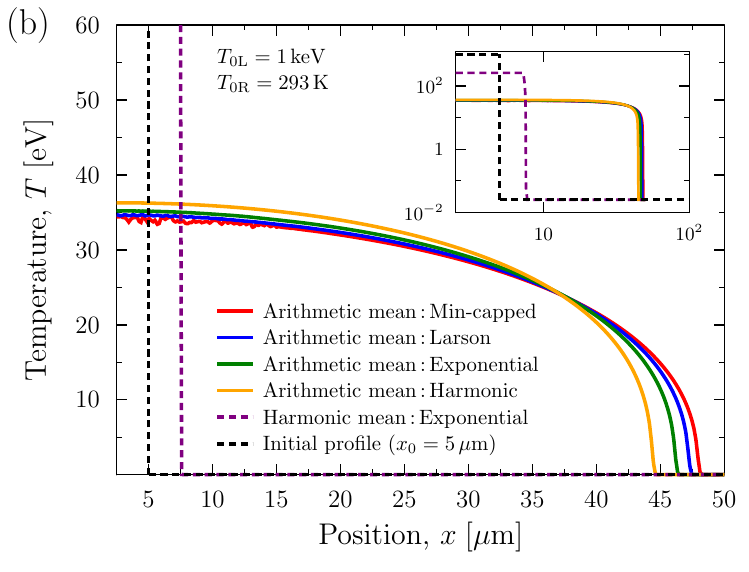}
		\label{fig:1d_thermal_wave_propagation_test}%
	}}
	\caption{
		(a): Effect of different interpolation functions $f_{\text{interp}}$ for applying flux-limitation to the conduction operator via an effective thermal conductivity, as discussed in the text. The dashed vertical line denotes the point where the unlimited (Fourier) heat flux \eqref{eq:fourier_heat_flux} is equal to the max heat flux \eqref{eq:max_heat_flux}. The min-capped (red curve) and harmonic (orange curve) represent bounding cases for the effective flux. (b): Spatial temperature profile for a simple nonlinear 1D thermal wave propagation test in an ideal gas (helium) of uniform density $\rho=0.01\;\unit{g\,cm^{-3}}$ with Spitzer-like thermal conductivity\cite{nrl_formulary}, comparing different options for the interpolation functions for apply flux-limitation and for evaluating quantities on the cell faces. The wave is initially strongly flux-limited, but settles to a diffusive solution everywhere except at the leading edge of the front after $t\approx0.2\;\unit{ns}$. Using the arithmetic mean \eqref{eq:arithmetic_mean} the wave quickly propagates from the hot region, as expected. The spread of heat front positions results from the different interpolation methods (color-coded with (a)). Using the harmonic mean \eqref{eq:harmonic_mean} (dashed purple curve), the heat front moves just over $2.5\;\unit{\mu m}$ from the initial position ($x_{0}=5\;\unit{\mu m}$) and maintains a temperature of over $250\;\unit{eV}$ behind the front. The inset in the top-right corner shows the profiles on a log-log scale to better emphasize the difference between the wave profiles using \eqa{eq:arithmetic_mean}{eq:harmonic_mean}.
	}
\end{figure*}

The ion heat flow is often either not flux-limited or its contribution is neglected entirely on the basis that electronic heat flow is usually dominant. The value of $\alpha_{i}$ is generally taken to be much larger than $\alpha_{e}$, although there is presently little-to-no consensus on a canonical value. Indeed, ionic VFP simulations strongly discredit the idea that a single value of $\alpha_{i}$ can universally describe ionic heat flow, especially in converging shocks \cite{Larroche_EuroPhysJourD_2003}. The default value used in this work is taken to be $0.5$ as this has been found in several other radiation-hydrodynamics codes, such as HYADES \cite{Larsen_JQSRT_1994} and DEIRA \cite{DEIRA_code_manual}. We consider the value of $\alpha_{i}$ to be one of the most uncertain aspects of our modeling.

\subsection{Interpolation functions}

Another important free choice in the implementation of flux-limited conduction is the manner in which the capping of the Fourier heat flow to the maximum value given by \eq{eq:max_heat_flux} is achieved. A particularly prevalent choice in the literature is to simply take the minimum value \cite{Atzeni_book, DEIRA_code_manual}
\begin{align}
\label{eq:min_capped_heat_flux}
Q_{\text{eff}}
= &\,
\min\rnd{Q_{\text{Four}}, Q_{\text{max}}}
\,,
\end{align}
which leaves the Fourier heat flow unmodified until the transition point. The impact of such a hard switch is discussed by Meezan et al.\cite{Meezan_PhysPlasmas_2020}, wherein the propagation of a nonlinear thermal wave is studied for $\alpha_{e} = 0.03-0.25$. The sharp front seen in their heat wave profiles for smaller flux-limiter coefficients results directly from the capping of the heat flux due to \eq{eq:min_capped_heat_flux}, which is unlikely to be representative of the true solution.

A smooth transition can of course be achieved in numerous ways, the most popular of which is to consider a harmonic interpolation \cite{Larsen_book, Ramis_ComputPhysComm_2009}
\begin{align}
\label{eq:harmonic_heat_flux}
Q_{\text{eff}}
= &\,
\rnd{1/Q_{\text{Four}} + 1/Q_{\text{max}}}^{\!-1}
\,.
\end{align}
Application of \eq{eq:harmonic_heat_flux} reduces the effective heat flux before the transition point (where $Q_{\text{Four}} = Q_{\text{max}}$), which removes the blunted thermal wave profile shape resulting from the hard switch to the floored value in \eq{eq:min_capped_heat_flux}. On the other hand, since the harmonic interpolation is a slow asymptotic approach to the maximum heat flux (\fig{fig:flux_limiter_interpolation_models}), this model will always retard the propagation thermal waves relative to using \eq{eq:min_capped_heat_flux}.

Less well-known smooth schemes that do not strongly alter the wave speed in the weakly flux-limited regime include the \quotes{Larson} model
\begin{align}
\label{eq:larson_heat_flux}
Q_{\text{eff}}
= &\,
\rnd{1/Q_{\text{Four}}^{\,2} + 1/Q_{\text{max}}^{\,2}}^{\!-1/2}
\,,
\end{align}
and the \quotes{exponential} model
\begin{align}
\label{eq:exponential_heat_flux}
Q_{\text{eff}}
= &\,
Q_{\mathrm{max}}
\sqr{1 - \exp\rnd{-Q_{\mathrm{Four.}}/Q_{\mathrm{max}}}}
\,.
\end{align}
In the Fourier limit, $Q_{\text{Four}} \ll Q_{\max}$, the exponential term in \eq{eq:exponential_heat_flux} can be Taylor expanded to leading order, whereupon one finds $Q_{\text{eff}} \to Q_{\text{Four}}$. In the flux-limited regime, $Q_{\text{Four}} \gg Q_{\max}$, the exponential term can simply be neglected to give $Q_{\text{eff}} \to Q_{\max}$. Both \eqa{eq:larson_heat_flux}{eq:exponential_heat_flux} originate in treatments of flux-limited radiation diffusion \cite{Mihalas_JQSRT_1982, Pomraning_book} and show similar behavior in how they approach $Q_{\text{max}}$ (\fig{fig:flux_limiter_interpolation_models}).

The electron and ion heat flux calculations are forced to follow Fourier's law \eqref{eq:fourier_heat_flux} in Hytrac, subsequently introducing a limited form of thermal conductivity \cite{DEIRA_code_manual, Ramis_ComputPhysComm_2009}
\begin{align}
\label{eq:maximum_thermal_conductivity}
\kappa_{\max} 
= &\,
Q_{\max}/\abs{\nabla T}
\,.
\end{align}
All the heat fluxes in the foregoing expressions \eqref{eq:min_capped_heat_flux}--\eqref{eq:exponential_heat_flux} can therefore be replaced with thermal conductivities, such that $\Qvec_{\text{eff}} = -\kappa_{\text{eff}}\nabla T$ by definition and 
\begin{align}
\label{eq:effective_thermal_conductivity}
\kappa_{\text{eff}}
= &\,
f_{\text{intrp}}\rnd{\kappa_{\text{tab}}, \kappa_{\max}}
\,,
\end{align}
where $\kappa_{\text{tab}}$ refers to the tabulated thermal conductivity at a given set of thermodynamic conditions and $f_{\text{interp}}$ represents one of the interpolation functions \eqs{eq:min_capped_heat_flux}{eq:exponential_heat_flux}. 

Allowing the limitation of the flux to enter via the thermal conductivity \eqref{eq:effective_thermal_conductivity} instead of directly acting on the magnitude of the heat flux has an additional benefit in the performance of Hytrac, which employs the Runge-Kutta-Legendre-2 explicit \quotes{super time-stepping} (STS) integration scheme for the conduction operators \cite{Meyer_JComputPhys_2014}. Specifically, by using \eq{eq:effective_thermal_conductivity} to define an effective thermal diffusivity coefficient $D_{\text{eff}} = \kappa_{\text{eff}}\rho/\tilde{C}_{V}$, one enforces consistency between the stability criterion for the diffusion time step, $\Delta t_{\text{dif}} = (\Delta x)^{2}/2D$, and the effective heat flux, $Q_{\text{eff}}$, in the flux-limited regime. Without this correction to the diffusion coefficient, the number of STS stages taken per hydrodynamic time step would be much larger than necessary. In the Fourier limit the regular STS scheme is of course exactly recovered.

\subsection{Evaluation of cell-face quantities}
\label{subsec:interpolation_to_cell_faces}

As with all codes based on the finite volume method, the heat fluxes between EGrid cells must be calculated on the cell faces and are evaluated via \cite{Blazek_book}
\begin{align}	
\label{eq:finite_volume_conducton_heating}
\left.\frac{\partial E}{\partial t}\right|_{\text{cond}}
= &\,
-\nabla \cdot \Qvec_{\text{eff}}
\stackrel{V \to 0}{=}
\frac{1}{V}\sum_{\cly{\text{faces}}} \rnd{\Svec\cdot\Qvec_{\text{eff}}}_{\text{face}}
\,,
\end{align}
in which $V$ is the cell volume, $\Svec$ is the cell face area vector and the summation extends over all cell faces. Thus, the effective thermal conductivity \eqref{eq:effective_thermal_conductivity} must be evaluated on the cell faces. Following \eq{eq:maximum_thermal_conductivity} the temperature gradient is approximated with a linear finite difference based on cell-centered values, e.g.\;$\abs{\nabla T}^{j+1/2} \approx \abs{T^{j+1} - T^{j}} / \delta$, where $\delta$ is the displacement between cell centers. 

As with the interpolation functions used to apply the limitation of the flux through the effective conductivity, there are several choices which can be made to evaluate the face-centered values of $\kappa_{\text{tab}}$ and $Q_{\max}$. This problem has previously received attention elsewhere for the standard form of the heat equation, i.e.\;as derived from Fourier's law \eqref{eq:fourier_heat_flux}, with results seeming to support different models for different situations \cite{Kadioglu_INL_2008}. The two most commonly-encountered models are the arithmetic mean (exemplified here using the tabulated thermal conductivity)
\begin{align}
\label{eq:arithmetic_mean}
\kappa_{\text{tab}}^{j+1/2}
= &\,
\frac{1}{2}\rnd{\kappa_{\text{tab}}^{j+1} + \kappa_{\text{tab}}^{j}}
\,,
\end{align}
and the harmonic mean
\begin{align}
\label{eq:harmonic_mean}
\kappa_{\text{tab}}^{j+1/2}
= &\,
2\frac{\kappa_{\text{tab}}^{j+1}\kappa_{\text{tab}}^{j}}
{\kappa_{\text{tab}}^{j+1} + \kappa_{\text{tab}}^{j}}
\,.
\end{align}
In \eqa{eq:arithmetic_mean}{eq:harmonic_mean} the superscripts refer to cell center (integer) and face (half-integer) indices. Other choices can be made to suit the specific nature of the problem \cite{Chang_MathComputModelling_1990}.

As discussed by Kadioglu et al. \cite{Kadioglu_INL_2008}, the arithmetic mean model can be formally derived from the steady state heat equation assuming that the thermal diffusivity is piecewise constant and continuous between (uniformly spaced) cell centers. It is the most natural choice when the thermodynamic fields in the system are smooth and continuous. The harmonic mean model is derived under similar assumptions, but asserts continuity of the diffusivities between the faces rather than the centers of the cells. It can also be obtained from continuity of the heat flux and temperature at the cell face and enforcing a discontinuity in the temperature gradient \cite{Tsui_NumHeatTrans_2014}. The harmonic mean model is popular in the heat transfer community \cite{Patankar_book}, especially for heterogeneous materials composed of layers with differing conductivities, where the heat flow is treated analogously to current flow through resistors in series. In particular, the harmonic mean approximately captures the contact resistance resulting from microscopic voids at the junction \cite{Incropera_book}. 

In the context of modeling heat flow in plasmas, an important and highly undesirable consequence of the harmonic mean is its failure to recover analytic results for well-known verification test cases \cite{Kamm_LANL_2008}. For example, the self-similarity solutions discussed by Zel'dovich and Raizer \cite{Zeldovich_book} and Reinicke and Meyer-ter-Vehn \cite{Reinicke_PhysFluidsA_1991}, both of which feature a nonlinear thermal wave propagating into a region with initially zero temperature (and therefore zero conductivity), are fundamentally incompatible with \eq{eq:harmonic_mean}. This is because the flux on the cell face between the heated and unheated parts of the domain evaluates to zero, thereby stalling wave propagation. 

Similar results are shown in \fig{fig:1d_thermal_wave_propagation_test}, which features an initially strongly flux-limited thermal wave propagating in helium of uniform density $\rho=0.01\;\unit{g\,cm^{-3}}$ with an initial temperature profile described by $T(x \leq x_{0},t=0) \equiv T_{0\text{L}} = 1\;\unit{keV}$ and $T(x > x_{0},t=0) \equiv T_{0\text{R}} = 2.5\times10^{-5}\;\unit{keV}\;(= 293\;\unit{K})$. Using the harmonic mean, the wave front (dashed purple curve) travels only $2.5\;\unit{\mu m}$ from the initial temperature discontinuity at $x_{0} = 5\,\unit{\mu m}$ compared to the $45-50\;\unit{\mu m}$ predicted using other model combinations. Note also the spread in the position of the wave front and the downstream temperature caused by varying the interpolation function $f_{\text{interp}}$. In other test cases, where the thermal wave propagates between regions with similar conductivities (e.g.\;Shafranov's self-similar two-temperature shock tube \cite{Shafranov_SovPhysJETP_1957}), results using \eq{eq:harmonic_mean} are extremely similar to those obtained using \eq{eq:arithmetic_mean}. Such a result is of course expected since in this case both methods should converge with sufficient resolution \cite{Maddix_JComputPhys_2018}.

Despite these concerns, we are motivated to include the harmonic mean option in our study due to its inclusion in other simulation codes used for HEDP research \cite{vanderHolst_AstrophysJSupplSeries_2011}. Furthermore, the choice of how to evaluate the maximum heat flux $Q_{\max}$ in \eq{eq:maximum_thermal_conductivity} at cell faces is also arbitrary and, in principle, independent to the choice applied to $\kappa_{\text{tab}}$. In this case, the same rationale behind the arithmetic mean interpolation of $Q_{\max}$ applies, although there is no clear analogue to the reasoning behind use of the harmonic mean approach. We consider the latter as a means of examining flux restriction under conditions where the heat flow is dominated by the free-streaming contribution \eqref{eq:max_heat_flux}.

\subsection{Conduction across a material interface}
\label{subsec:interface_conduction}

With the foregoing discussion in mind, one of the most important uncertainties in our modeling is the treatment of heat flow through material interfaces; especially for the deuterium-gold boundary, where the heated fuel is typically several orders of magnitude more conductive, and the heat flow could also be flux-limited. Using the arithmetic mean \eqref{eq:arithmetic_mean}, the properties of the material with the largest contribution (i.e.\;the largest value of $\kappa_{\text{tab}}$ or $Q_{\max}$) will dominate the flux, whereas using the harmonic mean \eqref{eq:harmonic_mean}, the converse is true. Considering heated fuel in contact with the initially cold anvil, this choice essentially governs if the interface is conducting (arithmetic) or insulating (harmonic). It will subsequently have a significant impact on the rate of conduction loss, since if the interface is treated as highly conductive, then the internal energy of the heated fuel will quickly leak into the high-heat capacity anvil and quench the contribution to the yield. The same concern also holds for the later stages, when the fuel is being compressed within the collapsing cavity. On the other hand, if the interface is strongly insulating, then the fuel ions will tend to stay hot through the main reverberation. Moreover, the evolution of the state is coupled to electronic conduction through equilibration, the rate of which is also influenced by the fact that the initial nonequilibrium produced by the shock is enhanced when ion conduction is suppressed.

Using Spitzer's well-known expressions for the thermal conductivities \cite{Spitzer_book} we may estimate the conditions for which the diffusive conduction across an interface between fully-ionized deuterium (D) and some high-Z material (Z) is driven by the conditions on one side only. Solving for the states that lead to a 10-fold disparity, i.e.\;$\kappa_{j}^{\text{D}} \geq 10\,\kappa_{j}^{\text{Z}}$, gives 
\allowdisplaybreaks
\begin{align}
\label{eq:kappa_e_interface_mismatch}
\frac{T_{e\,\text{D}}}{T_{e\,\text{Z}}}
\gtrsim &\,
0.6772\;\rnd{Z^{*}_{\text{Z}}}^{-2/5}
\,,
\\
\label{eq:kappa_i_interface_mismatch}
\frac{T_{i\,\text{D}}}{T_{i\,\text{Z}}}
\gtrsim &\,
2.512\;\rnd{Z_{\text{Z}}^{*}}^{-8/5}
\,\rnd{\frac{A_{\text{D}}}{A_{\text{Z}}}}^{1/5}
\,,
\end{align}
in which $A$ denotes the atomic mass and $Z_{\text{Z}}^{*}$ the effective ion charge for transport processes in the high-Z material. Note that the Coulomb logarithms have been assumed to be equal for simplicity. Taking $Z_{\text{Au}}^{*} \approx 6.4$ and $Z_{\text{PMMA}}^{*} \approx 2.9$ as representative ionization states around the time where the fusion output begins, \eqa{eq:kappa_e_interface_mismatch}{eq:kappa_i_interface_mismatch} suggest that the electron conductivity of the fuel dominates when $T_{e\,\text{D}} \gtrsim 0.113\;T_{e\,\text{Au}}$ and $T_{e\,\text{D}} \gtrsim 0.226\;T_{e\,\text{PMMA}}$. The equivalent criteria for the ions are $T_{i\,\text{D}} \gtrsim 0.0818\;T_{i\,\text{Au}}$ and $T_{i\,\text{D}} \gtrsim 0.425\;T_{i\,\text{PMMA}}$. These conditions are well-fulfilled for both the D-Au and D-PMMA interfaces in our simulations throughout the period of neutron emission.

Equivalently, in the strongly flux-limited regime one can solve for states with $Q_{\max\,j}^{\text{D}} \gtrsim 10\, Q_{\max\,j}^{\text{Z}}$, giving 
\begin{align}
\label{eq:Q_max_e_interface_mismatch}
\frac{T_{e\,\text{D}}}{T_{e\,\text{Z}}}
\gtrsim &\,
4.64 \, \rnd{\frac{n_{e\,\text{Z}}}{n_{e\,\text{D}}}}^{2/3}
\,,
\\
\label{eq:Q_max_i_interface_mismatch}
\frac{T_{i\,\text{D}}}{T_{i\,\text{Z}}}
\gtrsim &\,
4.64 \, \rnd{\frac{n_{i\,\text{Z}}}{n_{i\,\text{D}}}}^{2/3} \rnd{\frac{A_{\text{D}}}{A_{\text{Z}}}}^{1/3}
\,.
\end{align}
These conditions are much more restrictive than their diffusive regime counterparts \eqa{eq:kappa_e_interface_mismatch}{eq:kappa_i_interface_mismatch}, due to the weaker power of the temperature and the linear dependence of \eq{eq:max_heat_flux} on the number density. Again considering the conditions at the onset of neutron emission as an example, we have $\rho_{\text{D}} \approx 0.05\,\unit{g\,cm^{-3}}$, $\rho_{\text{Au}} \approx 11\,\unit{g\,cm^{-3}}$ and $\rho_{\text{PMMA}} \approx 6.5\,\unit{g\,cm^{-3}}$. From \eqa{eq:Q_max_e_interface_mismatch}{eq:Q_max_i_interface_mismatch} the following criteria for the heat flux to be dominated by the properties of the fuel are obtained: $T_{e\text{D}}\gtrsim 27.459\;T_{e\,\text{Au}}$ and $T_{e\text{D}}\gtrsim 106.46\;T_{e\,\text{PMMA}}$ for the electrons, and $T_{i\text{D}}\gtrsim 1.729\;T_{i\,\text{Au}}$ and $T_{i\text{D}}\gtrsim 35.954\;T_{i\,\text{PMMA}}$ for the ions. Of these, only the ionic transport across the interface between the fuel and the gold anvil is always well-fulfilled. Nevertheless, based on these simple estimates we expect a strong correlation between target performance and the choice of \eqa{eq:arithmetic_mean}{eq:harmonic_mean} for evaluating the heat fluxes.

It should, of course, be noted that these criteria are predicated on very simple estimates, especially with respect to the assumption of universality of $\alpha_{e}$ and $\alpha_{i}$ in the different materials, and to some degree obscure the scale and influence of the challenge regarding modeling interfacial transport. In particular, the coupling of different transport processes \cite{Haack_PhysRevE_2017} and the evolution of the interface itself in space and time \cite{Stanton_PhysRevX_2018} represent significant complications which go far beyond the scope of the present work.

\subsection{Uncertainties in conduction microphysics}
\label{subsec:microphysics_uncertainties}

\begin{figure*}
	\subfloat[]{
		\includegraphics[width=0.32\textwidth,trim={0.5cm 0.5cm 0.5cm 1cm},clip]
		{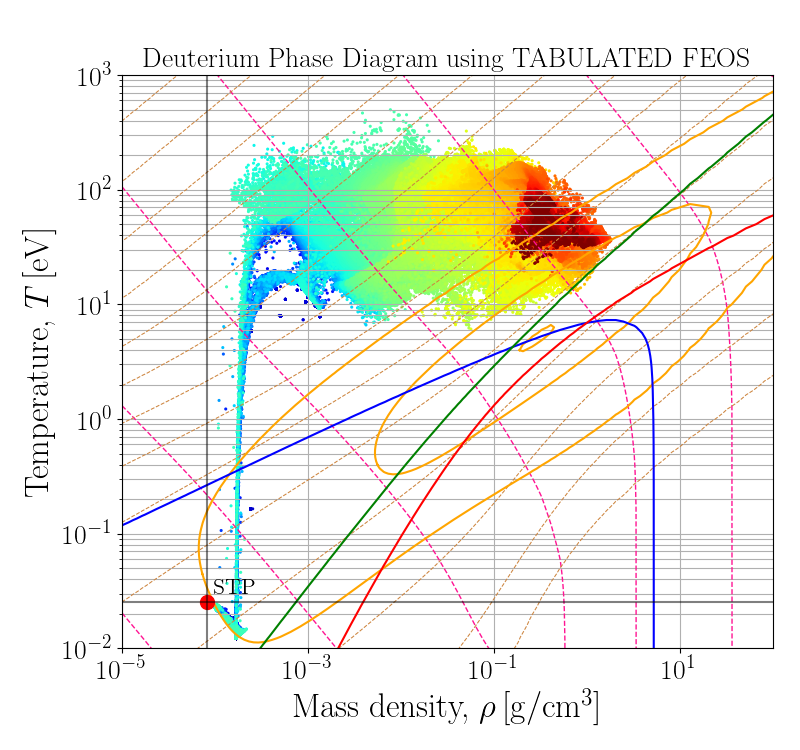}
		\label{fig:deuterium_cell_trajectories}
	}
	\subfloat[]{
		\includegraphics[width=0.32\textwidth,trim={0.5cm 0.5cm 0.5cm 1cm},clip]
		{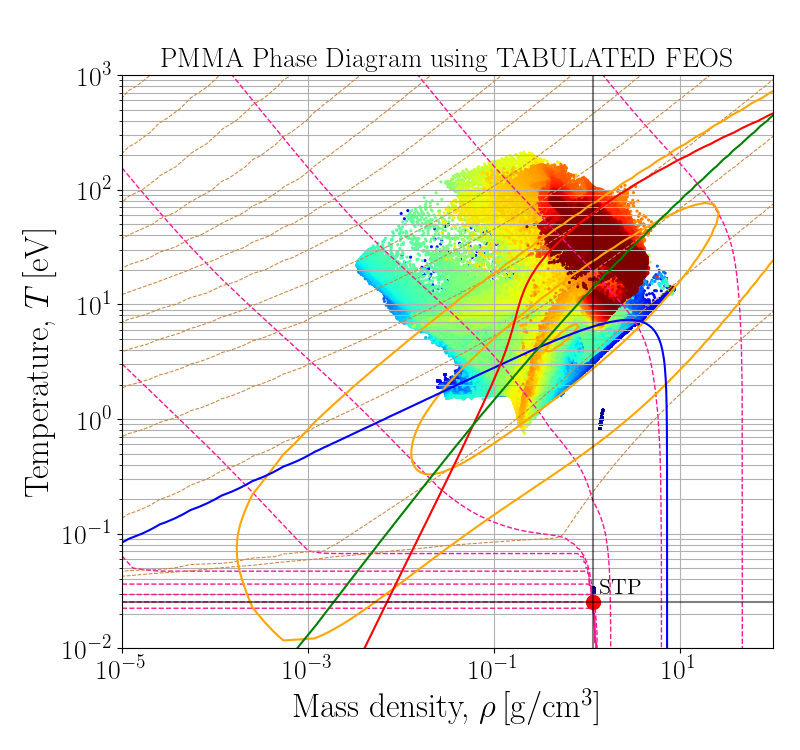}
		\label{fig:pmma_cell_trajectories}
	}
	\subfloat[]{
		\includegraphics[width=0.32\textwidth,trim={0.5cm 0.5cm 0.5cm 1cm},clip]
		{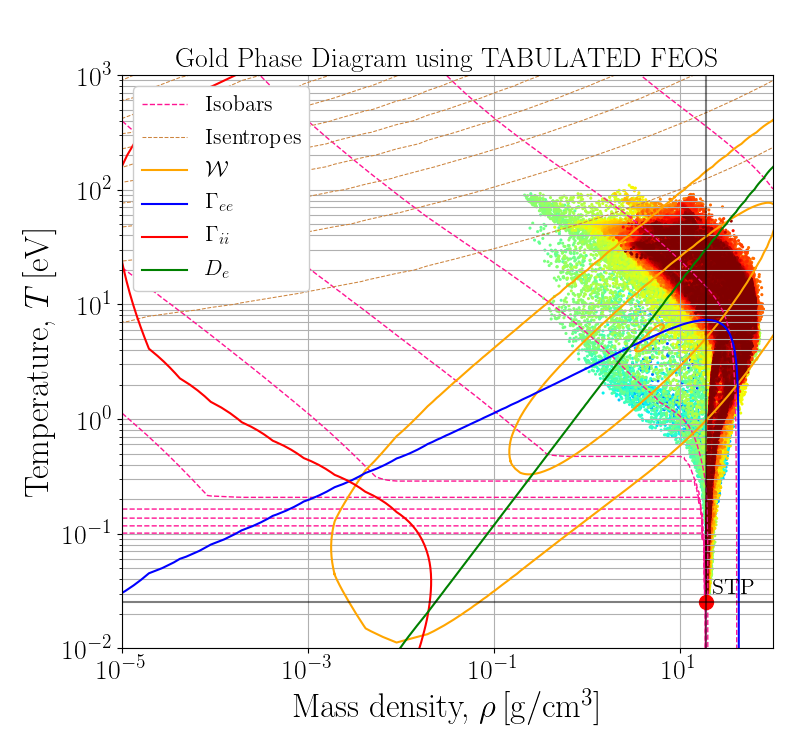}
		\label{fig:gold_cell_trajectories}
	}
	\caption{
		Cell trajectory plots on the $\rho-T_{i}$ plane for the fuel (a), coverslip (b) and anvil (c) materials. Contours for $\Gamma_{ii} = 1$ (red), $\Gamma_{e} = 1$ (blue), $D_{e} \equiv 8/(3\sqrt{\pi}\Theta^{3/2}) = 1$ (green) and $W = \cly{0.01, 0.1, 0.99}$ (orange) are shown. For the initially solid materials (PMMA and gold) the Maxwell construction region, where the accuracy of the FEOS model is highly uncertain, is denoted by the regions where the isobars abruptly become horizontal. There are no states in this region at any time in the simulation. The time evolution of the simulation is denoted by the color scale (blue - early time and dark red - late time). The initial conditions of the deuterium fuel are shifted from true STP conditions for $P_{0} = 1\;\unit{bar}$ due to the lack of molecular dissociation in FEOS. Ad hoc corrections following the linear mixing model \cite{Ross_JChemPhys_1983} have been found to make negligible difference to the BWA profiles or final neutron yield in these simulations, as expected for fuel initially in the gaseous phase.
	}
	\label{fig:cell_trajectories}
\end{figure*}

Aside from the details of the conduction operator implementation, the accuracy of the tabulated microphysics models which underpin the energetics must also be considered. In these simulations, we use the well-known Lee-More \cite{Lee_PhysFluids_1984} and Stanton-Murillo \cite{Stanton_PhysRevE_2016} models for the conductivities of the electrons and ions, respectively. The Lee-More model is corrected from the usual Lorentz approximation to include electron-electron scattering following the approach of Apfelbaum \cite{Apfelbaum_PhysPlasmas_2020}. For the electron-ion energy exchange rate, we use the $f$-sum rule approach \cite{Hazak_PhysRevE_2001, Gericke_JPhysConfSeries_2005}, which has been shown to perform well compared to molecular dynamics simulations \cite{Vorberger_HEDP_2014}. Higher-order considerations such as the coupled mode effect \cite{Dharma-wardana_PhysRevLett_1991, Dharma-wardana_PhysRevE_1998, Vorberger_PhysRevE_2010, Chapman_PhysRevE_2013} can safely be ignored since the nonequilibrium states produced in our simulations almost always have $T_{i} > T_{e}$. All the microphysics models are driven by the ionization predicted by FEOS for consistency with the EoS \cite{Faik_ComputPhysCommun_2018}.

Whilst each of these models include corrections for electron degeneracy and strong ion coupling, they are still not expected to be very accurate under conditions in the warm dense matter (WDM) regime. Of particular concern is the electron thermal conductivity as this is strongly influenced by many-body effects such as screening, structural effects and the ionization equilibrium of high-Z systems \cite{Ichimaru_PhysRevA_1985b, Kitamura_PhysRevE_1995, Hu_PhysRevE_2014b, Hu_PhysPlasmas_2016, Starrett_HEDP_2020, Shaffer_PhysRevE_2020}.

A simple but informative a priori indicator for model accuracy is encapsulated by the WDM parameter \cite{Murillo_PhysRevE_2010}
\begin{align}
\label{eq:murillo_wdm_parameter}
W(\rho,T_{e})
= &\,
S(\Theta)S(\Gamma_{ee}),~~
S(x) = 2/(x+1/x)
\,,
\end{align}
which provides a simple measure of the nonideality of thermodynamic states from the perspective of theoretical modeling. In \eq{eq:murillo_wdm_parameter}, the usual definitions of the degeneracy parameter, $\Theta$, and electron coupling parameter, $\Gamma_{ee}$, are used \cite{Murillo_PhysRevE_2010}. Cell trajectory plots extracted from the reference simulation (\fig{fig:deuterium_cell_trajectories}) show that the deuterium fuel is mostly ensconced within the ideal plasma regime, where $D_{e} \equiv 8/(3\sqrt{\pi}\Theta^{3/2}) \ll 1$ and $\Gamma_{ee,ii} \ll 1$, such that $W \ll 1$. The degree of nonideality in the fuel is seen to steadily increase at later times, where the neutron emission rate predicted by the reference case simulation is fastest, as the cavity collapse compresses the fuel whilst the temperature remains fairly constant. Significantly larger uncertainties can be expected for the properties of the coverslip (\fig{fig:pmma_cell_trajectories}) and anvil (\fig{fig:gold_cell_trajectories}) as they produce states with nonideal electrons, $W \sim 1$, and strongly coupled ions $\Gamma_{ii} \gtrsim 1$.

We account for the potential uncertainties in the microphysics by assessing the impact of a range of scaling factors, $s$, covering two orders of magnitude; $s = \cly{0.1, 0.5, 2, 10}$. These may be applied uniformly across phase space, referred to as a \quotes{blanket approach}, or conversely by a \quotes{targeted approach} which utilizes the WDM parameter \eq{eq:murillo_wdm_parameter} to focus their application toward regions with high uncertainty. The effective scaling parameter for a particular point in $\rho-T_{e}$ space is then
\begin{align}
\label{eq:targeted_scaling_factors}
s'(\rho,T_{e})
= &\,
1 + \rnd{s - 1}W\rnd{\rho,T_{e}}
\,.
\end{align}
The targeted form of the scaling factor crucially ensures that the tabulated microphysics are not unreasonably distorted under conditions where uncertainty in their forms is believed to be negligible, e.g.\;in the high-temperature, low-density limit where the well-established Spitzer-type models apply \cite{Spitzer_book}.

\subsection{Kinetic (nonlocal) effects}
\label{subsec:kinetic_effects}

\begin{figure}
	\subfloat{
		\includegraphics[width=0.49\textwidth]
		{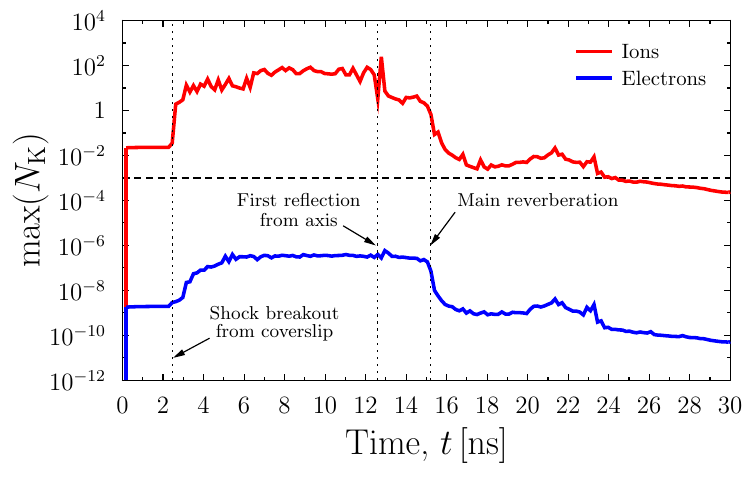}
	}
	\caption{
		Maximum ion and electron Knudsen numbers obtained over all fuel-containing cells as a function of time for the reference simulation discussed in Section \ref{subsec:sim_convergence}.
	}
	\label{fig:max_knudsen_number_histories}
\end{figure}
\begin{figure*}
	\subfloat[$t=15.56\,\unit{ns}$]{
		\includegraphics[width=0.32\textwidth,trim={0 0 0 3cm},clip]
		{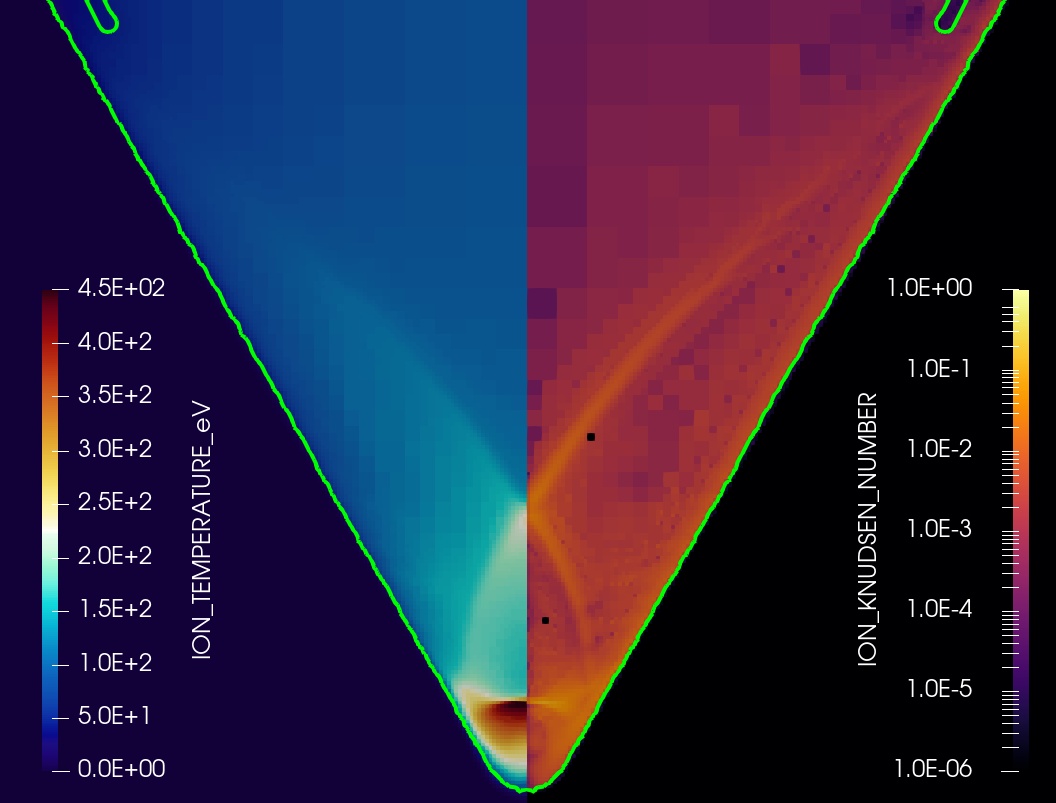}
		\label{fig:ion_knudsen_number_vs_ion_temperature_heat_map_t=15p56ns}
	}
	\subfloat[$t=18.03\,\unit{ns}$]{
		\includegraphics[width=0.32\textwidth,trim={0 0 0 3cm},clip]
		{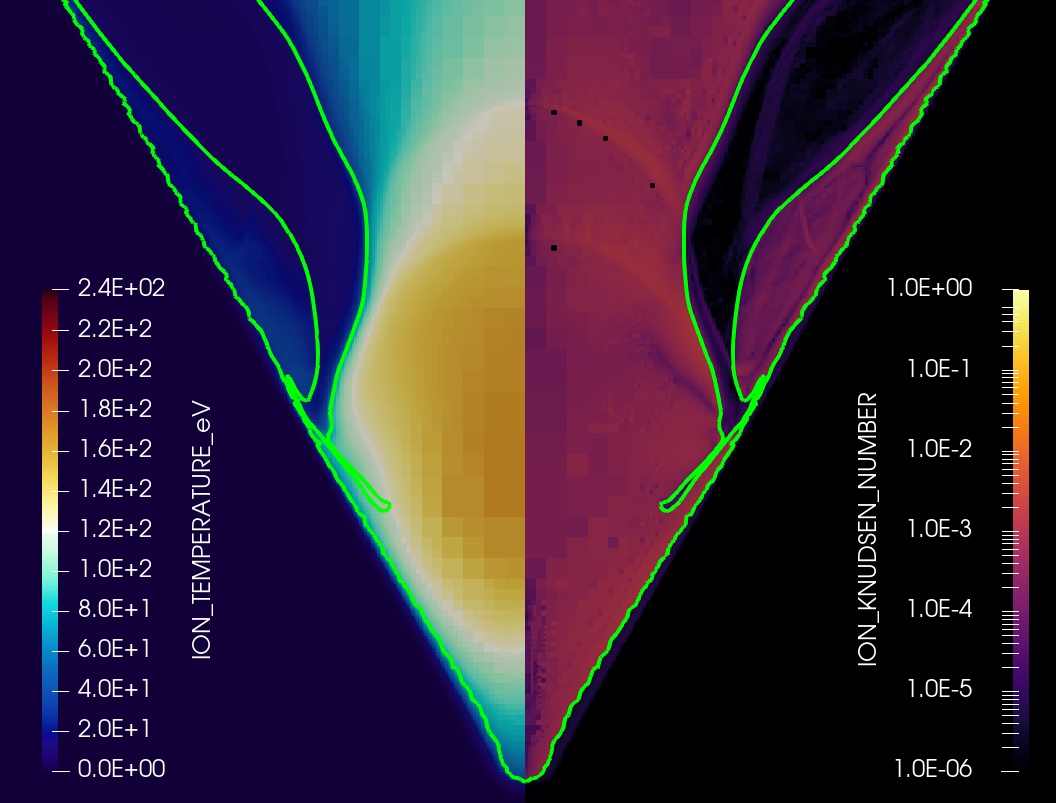}
		\label{fig:ion_knudsen_number_vs_ion_temperature_heat_map_t=18p03ns}
	}
	\subfloat[$t=20.13\,\unit{ns}$]{
		\includegraphics[width=0.32\textwidth,trim={0 0 0 3cm},clip]
		{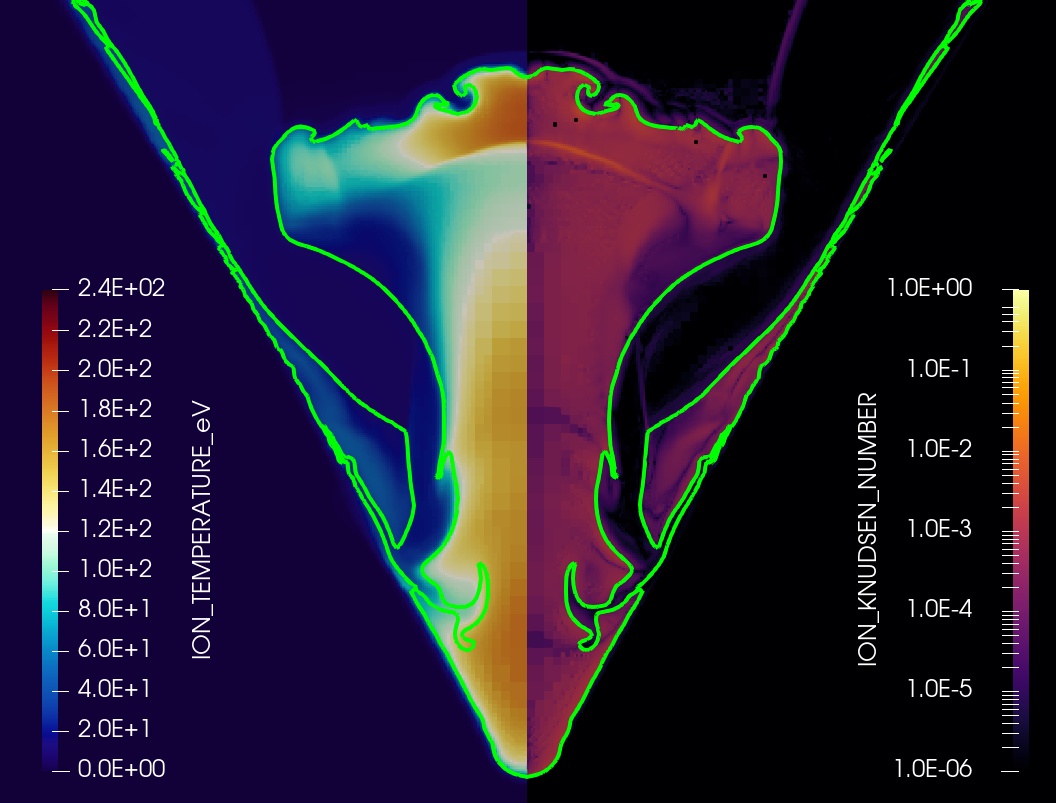}
		\label{fig:ion_knudsen_number_vs_ion_temperature_heat_map_t=20p13ns}
	}
	\caption{
		Heat maps of the ion temperature in units of $\unit{eV}$ (left-hand side) and ion Knudsen number (right-hand side) computed using \eqa{eq:Knudsen_number}{eq:plasma_gradient_scale_length} are shown immediately after the reverberation of the incident shock at the cavity tip (a) and at the beginning (b) and mid-way through (c) the compression phase. In these plots the bright green contour denotes the Lagrangian grid representing the material interface, which tracks the development of the PMMA jets expelled from the junction between the coverslip, anvil and fuel \cite{Charakhchyan_JourApplMechTechPhys_1994, Krasyuk_QuantElectronics_2005}.
	}
	\label{fig:knudsen_number_heat_maps}
\end{figure*}

A simple estimate of the degree of nonlocality with regard to collision-driven transport processes is the Knudsen number  
\begin{align}
\label{eq:Knudsen_number}
N_{\text{K}j}
= &\,
\frac{\lambda_{j}}{L}
\,.
\end{align}
Here the collisional mean free path, $\lambda_{j}$, of species $j = \cly{e,i}$ can be estimated from simple expressions \cite{Spitzer_book} throughout the simulation domain as a function of local thermodynamic variables. For the scale length, $L$, at a particular location in the system, we take the harmonic mean of inverse logarithmic derivatives
\begin{align}
\label{eq:plasma_gradient_scale_length}
L
\approx &\,
\rnd{\frac{1}{\sum_{X}}\sum_{X}\frac{\abs{\nabla X}}{X}}^{-1}
\,,
\end{align}
where in this case $X = \cly{T_{i}, T_{e}, \rho, Z_{i}^{*}}$ represents the set of variables which most strongly influence the collisional mean free path. This definition picks up both the change in density at a material interface and the strong gradient in heating at shock fronts and is similar to the scheme discussed by Taitano et al.\;\cite{Taitano_PhysPlasmas_2018}. For plasmas in which $N_{\text{K}} \gtrsim 10^{-3}$ the heat flux will be inhibited and a local, diffusive treatment is insufficient.

We find that the maximum electron Knudsen number is always negligibly small in the reference case simulation, being no larger $N_{\text{K}e} \lesssim 5\times10^{-7}$ (\fig{fig:max_knudsen_number_histories}). This strongly suggests that the electronic conduction should be well-described by purely local diffusion, i.e.\;using \eq{eq:fourier_heat_flux}. For the ions, very large maximum Knudsen numbers $N_{\text{K}i} \gtrsim 10$--$100$ are seen up to the first reflection from the simulation axis. Such high values are not unexpected during this phase as the incident shock is moving into ambient fuel, leading to very large temperature and density gradients at the front. We do not presently have any evidence to suggest they should be interpreted as being indicative of nonhydrodynamic flow. On the other hand, the values of order unity which occur during the main reverberation in close proximity to the D-Au interface (\fig{fig:knudsen_number_heat_maps}), which strongly suggests that the ionic transport into the anvil material will be nonlocal. Although values above the threshold of $N_{\text{K}i}=10^{-3}$ (horizontal dashed line in \fig{fig:max_knudsen_number_histories}) persist for much of the compression phase, these are generally still only around $N_{\text{K}i}\sim 10^{-2}$ and are restricted to the fronts of reflecting waves in the heated fuel. Moreover, the overall contribution of ionic conduction during this period is greatly reduced compared to electronic conduction as the fuel is essentially in equilibrium.

Finally, we note that another relevant consideration for plasmas characterized by large ion Knudsen numbers is the reduction of thermal reactivity \cite{Molvig_PhysRevLett_2012, Albright_PhysPlasmas_2013, Kagan_PhysRevLett_2015}. Work is presently being undertaken to examine the influence of this phenomenon and will be reported in a forthcoming publication.


\section{Results and discussion}
\label{sec:results}

To assess the impact of the configuration changes relative to the reference simulation (\tab{tab:reference_case}), we track three simple metrics which are sensitive to both the dynamics of the incident shock and the fuel energetics throughout the period of neutron emission:
\begin{enumerate}
	\item Time of first reflection of the incident shock from the axis of the simulation, $t_{\text{axis}}$
	\item Maximum BWA ion temperature produced in the main reverberation, $T_{\max}$.
	\item Mean value of BWA ion temperature during the compression phase, $T_{\text{av}}$.
\end{enumerate}
These are presented as percentage changes relative to the corresponding values obtained from the reference simulation, \;i.e. $\delta(M_{j}) = 100 \times (M_{j}/M_{j}^{\text{ref}} - 1)$, where $M_{j}$ stands for the relevant metric ($j = \{1,2,3\}$). We originally considered several other metrics, such as the full-width-half-maximum duration of the state produced in the main reverberation and the standard deviation of the BWA ion temperature in the compression phase; both of these proved unreliable for producing stable trends over the wide range of perturbations to the reference configuration. Most importantly, we have chosen not to track the total neutron yield resulting from each configuration change. This is principally because the plasma reactivity scales so strongly with ion temperature \cite{Atzeni_book} that any transient numerical artefacts present in the simulations can lead to spurious yields. Such artefacts are essentially impossible to correct for and distort otherwise meaningful and insightful trends; they are an unfortunate but inescapable consequence of the very low yields ($Y_{\text{n}} \gtrsim 100$) predicted for this target.

\begin{table}[b]
	\centering
	\begin{tabular}{c c}
		\hline\hline
		Variable & Value \\
		\hline
		Electron flux limiter interp. & Exponential \\
		Ion flux limiter interp. & Exponential \\
		Electron flux limiter coef. & 0.05 \\
		Ion flux limiter coef. & 0.5 \\
		Conductivity cell face interp. & Arithmetic mean \\
		Max heat flux cell face interp. & Arithmetic mean \\
		\hline
	\end{tabular}
	\caption{
		Configuration options of Hytrac's flux-limited conduction operator used in the reference simulation.
	}
	\label{tab:reference_case}
\end{table}

\begin{figure*}
	\includegraphics[width=\textwidth]
	{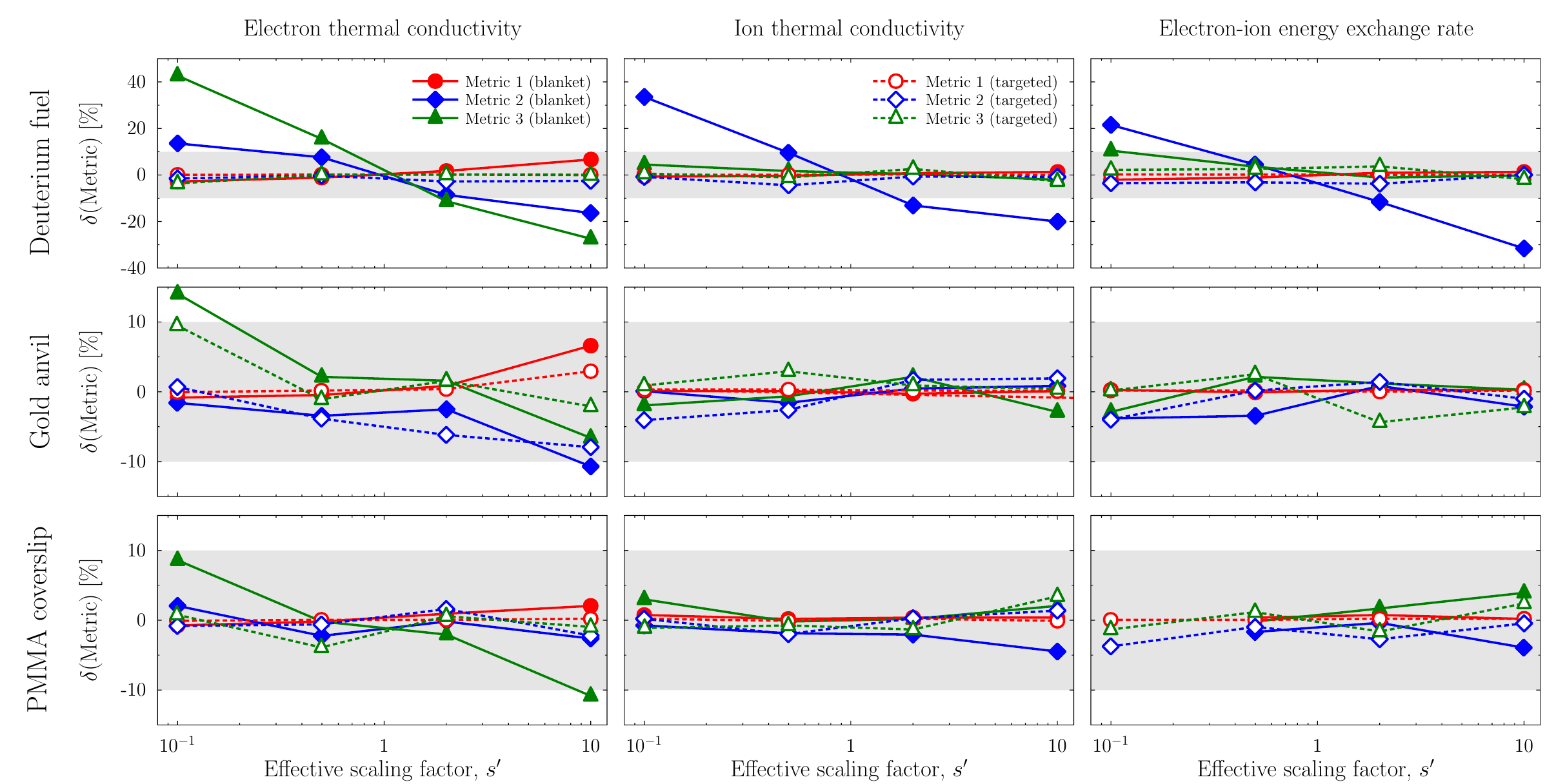}
	\caption{
		Effect of scaling factors applied on the conduction microphysics models featured the reference simulation; electron thermal conductivity (left-hand column), ion thermal conductivity (middle column) and electron-ion energy exchange rate (right-hand column). These are varied in the deuterium fuel (top row), gold anvil (middle row) and PMMA coverslip (bottom row). The three metrics described in the text are plotted in each panel as a percentage change relative to the reference case: Metric 1 (red markers/line); Metric 2 (blue markers/line); Metric 3 (green marker/line). The horizontal shaded band in each panel between $\delta = \pm 10\%$ represents the maximum impact relative to the reference simulation that we (arbitrarily) deem significant. In each panel the blanket (filled marker connected by solid lines) and targeted (open markers connected by dashed lines) approaches to applying the scaling factors are used.
	}
	\label{fig:scaling_factor_study_results}
\end{figure*}

We first examine the impact of the scaling factors applied to the components of the conduction microphysics; the electron and ion thermal conductivities and the electron-ion energy exchange rate. All three materials which directly contribute to the energetics of the collapsing cavity (the deuterium fuel, the PMMA coverslip and the gold anvil) are considered. For each material, scaling factors in the range $s = \cly{0.1, 0.5, 2, 10}$ have been applied (individually) using the \quotes{blanket} and \quotes{targeted} manner as described in Section \ref{subsec:microphysics_uncertainties}. To simplify the analysis we consider only order-one perturbations, i.e.\;only a single aspect of the simulation configuration of the reference case is perturbed at a time. This has the disadvantage that interactions between uncertainties are ignored, which is surely important in a two-temperature, conduction-dominated plasma due to the nonlinear feedback between the different energy loss mechanisms. On the other hand, establishing the basic trends is of prime importance and lays the foundations for more complex, coupled sensitivity studies in future.

Strong trends are found with respect to all the fuel microphysics using the blanket approach to applying the scaling factors (filled markers connected by solid lines in \fig{fig:scaling_factor_study_results}), particularly in the maximum BWA ion temperature achieved in the main reverberation (blue lines). This verifies the assertion that transport and equilibration play a crucial role in the energetics of the fuel during this phase of operation. However, the strongest correlation is between the electronic conductivity of the fuel and the average temperature in the compression phase (green lines), with substantially less sensitivity owing to the ionic conductivity and equilibration rate observed. Electron conduction is evidently the most important loss mechanism during this period, from which the majority of the neutron output is expected in the reference case. This can be readily understood since the plasma is essentially in thermal equilibrium during the cavity collapse, such that the relative importance of conduction is dictated by the ratio $\kappa_{e}/\kappa_{i} \sim (m_{i}/m_{e})^{1/2} \gg 1$. Weaker correlations are evident for the properties of the anvil and coverslip. The lack of any clear influence of any of the scaling factors on the arrival time of the first shock on the simulation axis (red lines) suggests that the early time dynamics of this target are mostly hydrodynamic in nature.

In all cases the sensitivity of the metrics to any of the scaling factors drops when the targeted approach \eqref{eq:targeted_scaling_factors} is used (open markers connected by dashed lines in \fig{fig:scaling_factor_study_results}). In this more realistic case none of the perturbations lead to a change beyond $\pm10\%$, which we arbitrarily define to constitute a \quotes{significant} deviation from the reference case. For the fuel this is unsurprising given the largely ideal states produced (\fig{fig:deuterium_cell_trajectories}). Although larger values of the scaling factor can be substantiated for the initially solid materials (the gold and PMMA) the results show that their thermophysical properties do not greatly influence target performance. The only exception is the possibility that the electron thermal conductivity of gold is underestimated by the LM model by a factor of ten in the WDM regime, where the simulation predicts the conditions remain throughout the duration of neutron production (\fig{fig:scaling_factor_study_results}). Although we have not been able to find any data in the literature to substantiate such a large discrepancy it could easily be accommodated by the general paucity of the LM model under conditions where structual properties limit the electron mean free path \cite{Lee_PhysFluids_1984, Ziman_AdvPhys_1967}. Firmer conclusions will only be able to be reached when more accurate thermal conductivity tables are available.

\begin{table*}
	\centering
	\begin{tabular}{c c c c c c}
		\hline\hline
		Config. option & Value & Eq. \# & $\delta(M_{1})\,[\%]$ & $\delta(M_{2})\,[\%]$ & $\delta(M_{3})\,[\%]$ \\
		\hline
		\multirow{3}{*}{Electron flux limiter interpolation method} & 
		Min. capped & \eq{eq:min_capped_heat_flux} & 
		0.644 & -3.097 & 1.723 \\
		 & 
		Harmonic & \eq{eq:harmonic_heat_flux} &
		-0.435 & 1.723 & 1.291 \\
		 & 
		Larson & \eq{eq:larson_heat_flux} &
		0.752 & -4.428 & 0.248 \\
		\hline
		\multirow{3}{*}{Ion flux limiter interpolation method} & 
		Min. capped & \eq{eq:min_capped_heat_flux} &
		0.212 & -9.140 & 1.795 \\
		 & 
		Harmonic & \eq{eq:harmonic_heat_flux} &
		0.284 & {\bf 12.155} & 4.025 \\
		 & 
		Larson & \eq{eq:larson_heat_flux} &
		0.248 & -4.212 & -0.183 \\
		\hline
		Conductivity cell face interpolation method & 
		Harmonic mean & \eq{eq:harmonic_mean} & 
		{\bf -10.317} & {\bf 118.333} & -1.666 \\
		\hline
		Max. heat flux cell face interpolation method &
		Harmonic mean & \eq{eq:harmonic_mean} &
		-3.968 & {\bf 179.666} & {\bf -10.317} \\
		\hline
	\end{tabular}
	\caption{
		Influence of changes to the various categorical, i.e. non-numerical, options in the conduction operator configuration. The metrics are expressed in terms of percentage changes relative to the reference simulation. The strongest changes result from changing the interpolation of the cell-centered thermal conductivities and max. heat flux values to the cell faces. 
	}
	\label{tab:conduction_operator_config changes}
\end{table*}

\begin{figure}
	\includegraphics[width=0.47\textwidth]
	{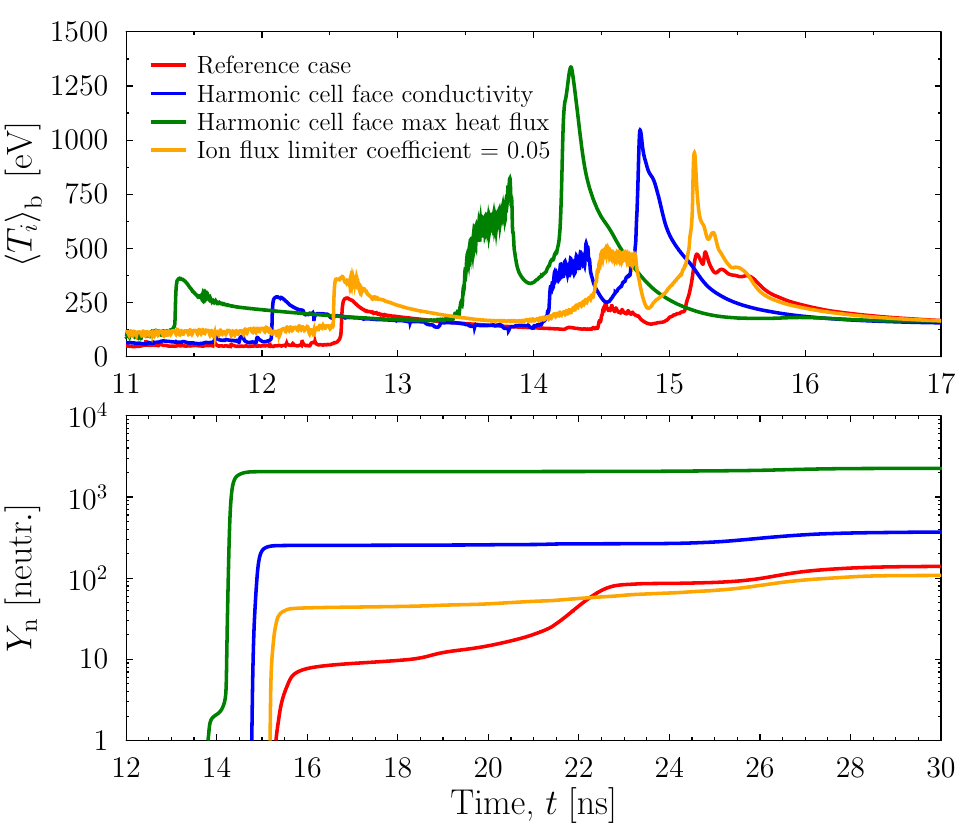}
	\caption{
		Comparison of the BWA ion temperature (top panel) and cumulative yield (bottom panel) from the reference case to those from simulations with the three most significant changes to the stagnation temperature. Note the change in the temporal scale on the $x$-axis between the top and bottom panels.
	}
	\label{fig:bwa_ion_temperature_and_cumulative_yield_comparison}
\end{figure}

Another factor which explains the lack of observed sensitivity to the properties of the anvil and coverslip is the dominant role of the fuel in determining the heat loss through the material interfaces. This can be understood since in the reference case the interfacial heat flux is constructed using the arithmetic mean method \eqref{eq:arithmetic_mean} and also because the conditions encapsulated by \eqs{eq:kappa_e_interface_mismatch}{eq:Q_max_i_interface_mismatch} are well-fulfilled in these simulations. Larger uncertainties consistent with the nonideal conditions in the anvil might therefore be expected if the interface is better described by a harmonic mean approximation, as in that case the properties of the cold, less conductive high-Z materials are more important in determining the heat flow out of the fuel. This has not been repeated here since the indications of the test cases discussed in Sec.\;\ref{subsec:interpolation_to_cell_faces} are that using \eq{eq:harmonic_mean} will not properly capture nonlinear thermal wave propagation through large conductivity gradients. Nevertheless, since we cannot rule the harmonic mean approach out without direct evidence, we are still inclined to compare its impact on the performance metrics.

As shown in \tab{tab:conduction_operator_config changes}, the impact of using the harmonic mean is indeed significant. In particular, the suppression of conduction loss from the fuel is seen to increase Metric 2 by a factor of two-to-three compared to the reference case, whereas the other metrics are much less strongly affected. With these changes the fusion output of the target occurs in a single \quotes{flash} of neutrons (\fig{fig:bwa_ion_temperature_and_cumulative_yield_comparison}), which constitutes almost all of the yield from the target. In contrast, the reference case simulation produces almost all of its yield during the compression phase. This fundamental change in the character of the neutron production is observed if either the unlimited (i.e.\;the tabulated data) or limited \eqref{eq:maximum_thermal_conductivity} components of the effective thermal conductivities are changed independently, with roughly $\sim30\%$ higher ion temperatures achieved when only the maximum heat flux is constructed on the cell face using \eq{eq:harmonic_mean}. The corresponding amplification in the yield is much larger, being roughy $\sim6$ times larger than for the case where only the tabulated conductivity is harmonically averaged. Taken together, these results are indicative of the ionic heat flow being moderately flux-limited, such that both contributions make substantial contributions to the total effective flux. 

Usage of the harmonic mean cell face interpolation method also leads to a faster initial shock, as shown by the earlier arrival time at the axis. This challenges the conclusion drawn from the scaling factor study: that conduction does not play a prominent role in the early-time dynamics. More work will be required to properly understand the role of conduction in this phase of operation. The fact that the average temperature during the compression phase is only weakly modified is readily understandable, however, since at this stage most of the conduction occurs within the fuel where temperature (and therefore conductivity) gradients are shallow.

\begin{figure}
	\includegraphics[width=0.47\textwidth]
	{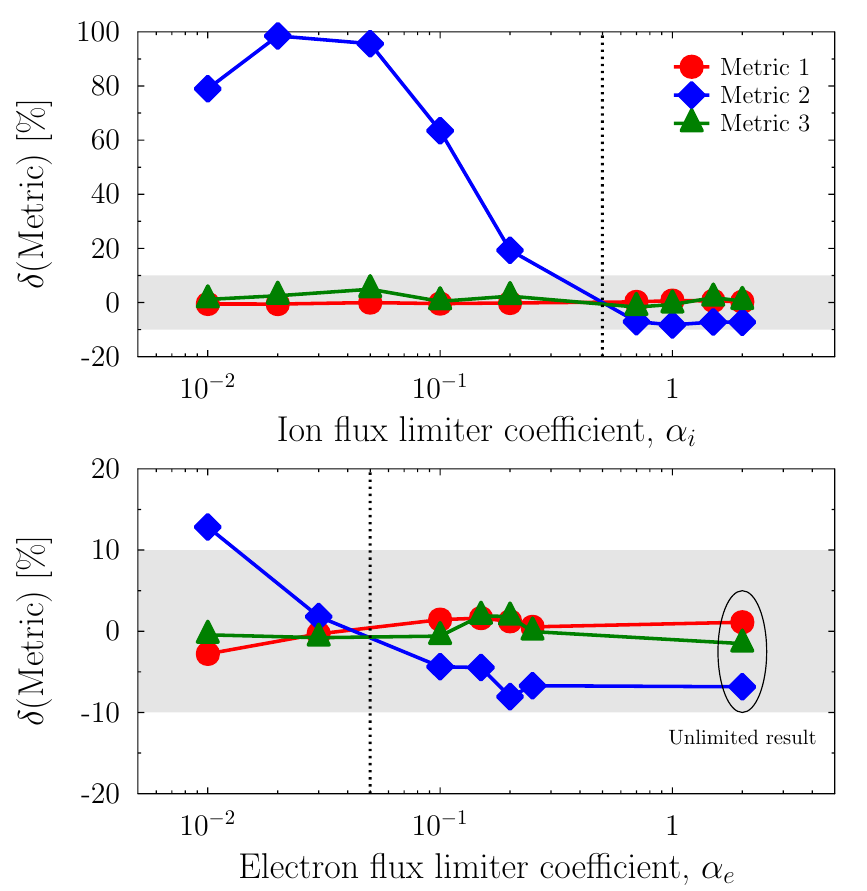}
	\caption{
		Variation of the same sensitivity metrics shown in \fig{fig:scaling_factor_study_results}, but considering the influence of flux limiter coefficients of the ions (top) and electrons (bottom). Note that the values associated with $\alpha_{e} = 2$ are actually those obtained for the case where the electron heat flow is purely diffusive, which is achieved by setting the interpolation function to $f_{\text{inter}}(\kappa_{\text{tab}}, \kappa_{\max}) = \kappa_{\text{tab}}$. The dotted vertical lines indicate the values from the reference case.
	}
	\label{fig:flux_limiter_coefficients_study_results}
\end{figure}

Interestingly, the target operation is found to be similarly affected without using the harmonic mean if ionic conduction is instead suppressed through the flux limiter coefficient. Setting $\alpha_{i} = \alpha_{e} = 0.05$, which amounts to a ten-fold reduction in the maximum ion heat flux from the reference simulations, again leads to a reverberation temperature nearly twice as hot as the reference case. On the other hand, the other metrics are very weakly affected. This is consistent with the findings from the scaling factor study where the role of ion conduction during the compression phase was shown to be negligible (\fig{fig:scaling_factor_study_results}). The fact that there is very little change to any of the metrics if the fraction of the maximum ion heat flux is allowed to increase arbitrarily, i.e.\;for $\alpha_{i} \geq 1$, again suggests that the ionic heat flow is not strongly flux-limited, but rather can be considered \quotes{quasi-nonlocal}. This is further supported by the observation that switching from the exponential interpolation method \eqref{eq:exponential_heat_flux} used in the reference case to the softer harmonic interpolation \eqref{eq:harmonic_heat_flux} leads to a slightly hotter main reverberation by cutting the effective heat flux; using the min-capped approach \eqref{eq:min_capped_heat_flux} slightly cools this feature by almost the same degree. The ratio $Q_{\text{eff}}/Q_{\max}$ must therefore be of order unity, where the difference between the min-capped and harmonic methods to the exponential model are roughly equal and opposite (see \fig{fig:flux_limiter_interpolation_models}). In the strongly flux-limited regime one would expect a much smaller effect from switching from exponential to min-capped interpolation. 

In contrast to the ions, the evidence does not suggest that the electronic heat flow is likely to be flux-limited, as no significant sensitivities are observed for any of the configuration changes considered with $\alpha_{e} > 0.05$. This is in agreement with the small Knudsen numbers noted in Section \ref{subsec:kinetic_effects}. We have validated this result by re-running the reference case simulation without any flux limitation applied to the electrons, which produced almost identical results to the case with $\alpha_{e} = 0.25$ (see the circled point placed at $\alpha_{e} = 2$ for convenience in \fig{fig:flux_limiter_interpolation_models}). Whilst a small increase in the peak temperature in the main reverberation can be obtained by decreasing the electron flux limiter coefficient to $\alpha_{e}=0.01$, the impact is far less than for the ions. Furthermore, none of the literature examined in this work support such small values.

It is important to note that only separate perturbations to the operator configuration have been examined in this work. This means that the potential for large impacts on performance caused by coupling of multiple configuration changes should be assessed before stronger conclusions can be drawn. From the perspective of being able to quantitatively compare predicted and experimental yields, there is certainly scope for thermal conduction to be the dominant factor in explaining the observed difference to Ref.\;\cite{Derentowicz_JourTechPhys_1977b}. Nevertheless, we note that none of the changes considered in this work resulted in a total yield close to the lowest reported yields of Derentowicz et al.; the highest total yield being roughly five times smaller. The work required to bring together all these threads in a properly coupled manner is presently ongoing, as are efforts to develop our understanding of the possible role and impact of kinetic ion transport on these targets and the influence of the other radiation-hydrodynamic phenomena left out of this work.


\section{Conclusions}
\label{sec:conclusions}

We have undertaken an initial examination of the influence of flux-limited thermal conduction modeling on the robustness and performance of the uniaxially-driven conical fusion targets used in the experiments of Derentowicz et al. \cite{Derentowicz_JourTechPhys_1977,Derentowicz_JourTechPhys_1977b}. This kind of target is of interest to FLF as it is one of the few examples of fusion targets in the literature with observable neutron yield, where the collapse is driven by a single planar shock produced by conditions within reach of our current experimental facilities. The attention given to thermal conduction in this paper is justified since other phenomena such as radiation transport and viscous effects are not expected to play as important a role in determining the energetics of the fuel. 

The uncertainties which principally affect conduction loss are associated with the related plasma microphysics and the degree to which the transport is nonlocal. To assess the impact of uncertainties in the microphysics we have undertaken scaling factor studies in which the tabulated models are scaled both uniformly, i.e.\;a blanket approach, and with a more realistic method using the WDM parameter \cite{Murillo_PhysRevE_2010} to target regions of density-temperature space with larger theoretical uncertainty. The reference case was produced using an idealized model that was rigorously converged on the basis of root-mean-square differences between time-dependent profiles of burn-weighted average (BWA) plasma parameters from successively higher resolution simulations. 

Three simple metrics which capture the dynamics of the initial stages of the cavity collapse and the energetics of the fusion fuel over the time of neutron emission were defined, from which direct comparison can be made between different simulations and the reference case. Uniform application of the scaling factors showed that electron conduction in the fuel during the compression phase most strongly affects the target performance. With the more realistic targeted approach we found very little influence on any of the metrics. This is due to the fact that it is the properties of the fuel that determine its energetics in the compression phase, which remains strongly ideal, and therefore minimally uncertain, throughout the simulations. We found that the role of the nonideal states expected to be produced in the plastic coverslip and gold anvil is minimised due to the numerical treatment of the heat flux at material interfaces (arithmetic averaging of transport coefficients on cell faces).

Considering alternative interpolation schemes for evaluating cell-centered quantities at cell faces leads to significant changes to the predicted conditions in the target and also in the character of the neutron emission. In particular, we found that the largest influence our performance metrics comes from changing the cell-face averaging from the arithmetic mean used in the reference case to a harmonic mean. This option is often favoured in the heat transfer community for heterogeneous materials, but does not perform well in well-known simple thermal wave propagation tests, in which it fails to correctly propagate nonlinear thermal waves between regions of substantially different conductivities. Without direct validation, however, this model cannot yet be ruled out completely. In the context of our simulations its main influence to produce a hotter reverberation state at the cavity tip. Specifically, the largest increase was observed when the harmonic mean was applied to the flux-limited component of the effective conductivity. In both cases, however, the fusion output is dominated by a single neutron flash produced almost entirely during the main reverberation, in contrast to the steady emission through the compression phase predicted by the reference simulation.

Similar changes in the performance metrics and target operation were found without using the harmonic mean by restricting the ionic heat flux in the reference case through the ion flux limiter coefficient. Reducing the default value from $\alpha_{i} = 0.5$ to $\alpha_{i} \leq 0.05$ was shown to also result in a hotter main reverberation state and consequently a strong single neutron flash. Increasing the value of $\alpha_{i}$ was found to have a much smaller impact, suggesting that the ionic heat flow is likely to be moderately flux-limited, i.e.\;nonlocal kinetic effects may be important, but not dominant. On the other hand, the electronic heat flow was shown to be well-approximated with a purely diffusive treatment, with the only exception being for the case where associated flux limiter coefficient was strongly reduced relative to the reference case, e.g.\;$\alpha_{e} \sim 0.01$. Since such small values are supported neither by experiments nor advanced transport simulations, we conclude that the uncertainty of our modelling to electronic heat flow is low. 

Assuming that the heat flow in the target is indeed well-described with the arithmetic mean for the evaluation of the fluxes, and ionic flux limiter coefficients much larger than those commonly used for the electrons, then our reference case simulations should be quite robust. The question of why the predicted total yield of these targets is several orders of magnitude less than those reported in Ref.\;\cite{Derentowicz_JourTechPhys_1977} therefore remains an open one. Moreover, we noted that none of the singular changes to the reference case considered in this work amplified the neutron yield by more than an order of magnitude. Whether or not this gap can be closed by multiple simultaneous changes and/or including a broader ranges of physics models (e.g.\;radiation transport and viscous effects) will require extensive multivariate sensitivity studies. We are presently investigating all of these factors as part of our ongoing investigations into this platform and will report on more findings in future publications.

Finally, we note that the findings related to the potentially important role of ionic conduction are interesting as they suggest that targets similar to those studied here may provide an experimental platform for accessing states of matter in the quasi-nonlocal regime, in which energy transport is close to the transition between local diffusion and kinetic transport. With suitable diagnostics to provide time-resolved histories of neutron production and electron and ion temperatures from the main reverberation state, it may be possible to experimentally distinguish between predictions based on simple ion transport models (such as flux-limited conduction) and state-of-the-art kinetic models, whilst minimizing uncertainty associated with electron transport and equilibration. Such data may prove to be invaluable for benchmarking integrated simulation codes for ICF.


\section*{Data availability}

The data that support the findings of this study are available from the corresponding author upon reasonable request.


\section*{Acknowledgements}

The authors gratefully acknowledge R. King and J. Gardiner for providing HPC support and A. Venskus and J. Herring for invaluable contributions to the robustness and stability of Hytrac. We also recognize insightful conversations with A. Crilly and G. Kagan (Imperial College London) regarding equation of state and ionic transport modeling. Finally, the authors would like to thank the referees for offering invaluable scrutiny and insights during the review process, which have helped to enhance the clarify and quality of the manuscript.


\bibliographystyle{unsrt}
\bibliography{books,papers,phd_theses,misc}

\begin{thebibliography}{10}

\bibitem{Gaffney_NuclFusion_2013}
J.~A. Gaffney, D.~Clark, V.~Sonnad, and S.~B. Libby.
\newblock {Development of a Bayesian method for the analysis of inertial
  confinement fusion experiments on the NIF}.
\newblock {\em Nuclear Fusion}, 53:073032, 2013.

\bibitem{Gaffney_HEDP_2013}
J.~A. Gaffney, D.~Clark, V.~Sonnad, and S.~B. Libby.
\newblock {Bayesian inference of inaccurancies in radiation transport physics
  from inertial confinement fusion experiments}.
\newblock {\em High Energy Density Physics}, 9:457, 2013.

\bibitem{Melvin_PhysPlasmas_2015}
J.~Melvin, H.~Lim, V.~Rana, B.~Cheng, J.~Glimm, D.~H. Sharp, and D.~C. Wilson.
\newblock {Sensitivity of inertial confinement fusion hot spot properties to
  the deuterium-tritium fuel adiabat}.
\newblock {\em Physics of Plasmas}, 22:022708, 2015.

\bibitem{Guymer_PhysPlasmas_2015}
T.~M. Guymer, A.~S. Moore, J.~Morton, J.~L. Kline, S.~Allan, N.~Bazin,
  J.~Benstead, C.~Bentley, A.~J. Comley, J.~Cowan, K.~Flippo, W.~Garbett,
  C.~Hamilton, N.~E. Lanier, K.~Mussack, K.~Obrey, L.~Reed, D.~W. Schmidt,
  R.~M. Stevenson, J.~M. Taccetti, and J.~Workman.
\newblock {Quantifying equation-of-state and opacity errors using integrated
  supersonic diffusive radiation flow experiments on the National Ignition
  Facility}.
\newblock {\em Physics of Plasmas}, 22:043303, 2015.

\bibitem{Derentowicz_JourTechPhys_1977}
H.~Derentowicz and S.~Kaliski aand Z.~Zi\'{o}{\l}kowski.
\newblock {Generation of fusion neutrons in a deuterium filled cone by means of
  explosive implosion of polythene shell. Part 1. Theoretical estimations}.
\newblock {\em Journal of Technical Physics}, 18:465--471, 1977.

\bibitem{Derentowicz_JourTechPhys_1977b}
H.~Derentowicz, S.~Kaliski, J.~Wolski, and Z.~Zi\'{o}{\l}kowski.
\newblock {Generation of thermonuclear fusion neutrons by means of a pure
  explosion. Part 2. Experimental results}.
\newblock {\em Journal of Technical Physics}, 25:135--147, 1977.

\bibitem{Krasyuk_QuantElectronics_2005}
I.~K. Krasyuk, A.~Yu. Semenov, and A.~A. {Charakhch'yan}.
\newblock {Use of conic targets in inertial confinement fusion}.
\newblock {\em Quantum Electronics}, 35(9):769, 2005.

\bibitem{Shyam_AtomkernenergieKerntechnik_1984}
A.~Shyam and M.~Srinivasan.
\newblock {Investigation of the feasibility of conical target compression
  experiments using exploding foil driven hyper-velocity liners}.
\newblock {\em Atomkernenergie Kerntechnik}, 44:196, 1984.

\bibitem{Vovchenko_JETPLett_1977}
V.~I. Vovchenko, A.~S. Goncharov, Iu.~S. Kasianov, O.~V. Kozlov, I.~K. Krasiuk,
  A.~A. Maliutin, M.~G. Pastukhov, P.~P. Pashinin, and A.~M. Prokhorov.
\newblock {Generation of thermonuclear neutrons by laser action on a conical
  target}.
\newblock {\em Journal of Experimental and Theoretical Physics Letters},
  26:628, 1977.

\bibitem{Mason_ApplPhysLett_1979}
R.~J. Mason, R.~J. Fries, and E.~H. Farnum.
\newblock {Conical targets for implosion studies with a $\mathrm{CO}_{2}$
  laser}.
\newblock {\em Applied Physics letters}, 34:14, 1979.

\bibitem{Bogolyubskii_JETPLett_1976}
S.~L. Bogolyubskii, B.~P. Gerasimov, V.~I. Liksonov, A.~P. Mikhailov,
  Y.~P.Popov, L.~I. Rudakov, A.~A. Samarskii, and V.~P. Smirnov.
\newblock {Thermonuclear neutron yield from a plasma compressed by a shell}.
\newblock {\em Journal of Experimental and Theoretical Physics Letters},
  24:182, 1976.

\bibitem{Spitzer_PhysRev_1953}
L.~Spitzer and R.~H\"{a}rm.
\newblock {Transport phenomena in a completely ionized gas}.
\newblock {\em Physical Review}, 89:977, 1953.

\bibitem{Ferziger_book}
J.~H. Ferziger and H.~G. Kaper.
\newblock {\em {Mathematical Theory of Transport Processes in Gases}}.
\newblock North-Holland, 1972.

\bibitem{Larsen_book}
Jon Larsen.
\newblock {\em {Foundations of High-Energy-Density Physics}}.
\newblock Cambridge University Press, 2017.

\bibitem{Graziani_book}
F.~Graziani, M.~P. Dejarlais, R.~Redmer, and S.~B. Trickey.
\newblock {\em {Frontiers and Challenges in Warm Dense Matter: Lecture Notes in
  Computational Science and Engineering}}.
\newblock Springer International Publishing Switzerland, 2014.

\bibitem{Larroche_EuroPhysJourD_2003}
O.~Larroche.
\newblock {Kinetic simulations of fuel ion transport in ICF target implosions}.
\newblock {\em European Physics Journal D}, 27:131, 2003.

\bibitem{Larroche_PhysPlasmas_2012}
O.~Larroche.
\newblock {Ion Fokker-Planck simulation of $\mathrm{D}-{}^{3}\mathrm{He}$ gas
  target implosions}.
\newblock {\em Physics of Plasmas}, 19:122706, 2012.

\bibitem{Taitano_JourComputPhys_2018}
W.~T. Taitano, L.~Chac\'{o}n, and A.~N. Simakov.
\newblock {An adaptive, implicit, conservative, $1D-2V$ multi-species
  Vlasov-Fokker-Planck multi-scale solver in planar geometry}.
\newblock {\em Journal of Computational Physics}, 365:173, 2018.

\bibitem{Rinderknecht_PlasmaPhysControlFusion_2018}
H.~G. Rinderknecht, P.~A. Amendt, S.~C. Wilks, and G.~Collins.
\newblock {Kinetic physics in ICF: presentunderstanding and future directions}.
\newblock {\em Plasma Physics and Controlled Fusion}, 60:064001, 2018.

\bibitem{Schurtz_PhysPlasmas_2000}
G.~P. Schurtz, Ph.~D. Nicola\"{i}, and M.~Busquet.
\newblock {A nonlocal electron conduction model for multidimensional radiation
  hydrodynamics codes}.
\newblock {\em Physics of Plasmas}, 7:4238, 2000.

\bibitem{Manheimer_PhysPlasmas_2008}
W.~Manheimer, D.~Colombant, and V.~Goncharov.
\newblock {The development of a Krook model for nonlocal transport in laser
  produced plasmas. I. Basic theory}.
\newblock {\em Physics of Plasmas}, 15:083103, 2008.

\bibitem{Colombant_PhysPlasmas_2008}
D.~Colombant and W.~Manheimer.
\newblock {The development of a Krook model for nonlocal transport in laser
  produced plasmas. II. Application of the theory and comparisons with other
  models}.
\newblock {\em Physics of Plasmas}, 15:083104, 2008.

\bibitem{Cao_PhysPlasmas_2015}
D.~Cao, G.~Moses, and J.~Delettrez.
\newblock {Improved non-local electron thermal transport model for
  two-dimensional radiation hydrodynamics}.
\newblock {\em Physics of Plasmas}, 22:082308, 2015.

\bibitem{DelSorbo_PhysPlasmas_2015}
D.~{Del Sorbo}, {J.-L.} Feugeas, Ph. Nicola\'{ı}, M.~{Olazabal-Loume},
  B.~Dubroca, S.~Guisset, M.~Touati, and V.~Tikhonchuk.
\newblock {Reduced entropic model for studies of multidimensional nonlocal
  transport in high-energy-density plasmas}.
\newblock {\em Physics of Plasmas}, 22:082706, 2015.

\bibitem{Holec_PhysPlasmas_2018}
M.~Holec, J.~Nikl, and S.~Weber.
\newblock {Nonlocal transport hydrodynamic model forlaser heated plasmas}.
\newblock {\em Physics of Plasmas}, 25:032704, 2018.

\bibitem{Hoffman_PhysPlasmas_2015}
N.~M. Hoffman, G.~B. Zimmerman, K.~Molvig, H.~G. Rinderknecht, M.~J. Rosenberg,
  B.~J. Albright, A.~N. Simakov, H.~Sio, A.~B. Zylstra, M.~Gatu Johnson, F.~H.
  S\'{e}guin, J.~A. Frenje, C.~K. Li. R.~D. Petrasso, D.~M. Higdon,
  G.~Srinivasan, V.~Yu. Glebov, C.~Stoeckl, W.~Seka, and T.~C. Sangster.
\newblock {Approximate model for the ion-knetic regime in
  inertial-confinement-fusion capsule implosions}.
\newblock {\em Phyics of Plasmas}, 22:052707, 2015.

\bibitem{Brodrick_PhysPlasmas_2017}
J.~P. Brodrick, R.~J. Kingham, M.~M. Marinak, M.~V Patel, A.~V. Chankin, J.~T.
  Omotani M.~V. Umansky, D.~{Del Sorbo}, B.~Dudson, J.~T. Parker, G.~D. Kerbel,
  M.~Sherlock, and C.~P. Ridgers.
\newblock {Testing nonlocal models of electron thermal conduction for magnetic
  and inertial confinement fusion applications}.
\newblock {\em Physics of Plasmas}, 24:092309, 2017.

\bibitem{Sherlock_PhysPlasmas_2017}
M.~Sherlock, J.~P. Brodrick, and C.~P. Ridgers.
\newblock {A comparison of non-local electron transport model for laser-plasmas
  relevant to inertial confinement fusion}.
\newblock {\em Physics of Plasmas}, 24:082706, 2017.

\bibitem{Henchen_PhysRevLett_2018}
R.~J. Henchen, M.~Sherlock, W.~Rozmus, J.~Katz, D.~Cao, J.~P. Palastro, and
  D.~H. Froula.
\newblock {Observation of Nonlocal Heat Flux Using Thomson Scattering}.
\newblock {\em Physical Review Letters}, 121:125001, 2018.

\bibitem{Henchen_PhysPlasmas_2019}
R.~J. Henchen, M.~Sherlock, W.~Rozmus, J.~Katz, P.~E. {Masson-Laborde}, D.~Cao,
  J.~P. Palastro, and D.~H. Froula.
\newblock {Measuring heat flux from collective Thomson scattering with
  non-Maxwellian distribution functions}.
\newblock {\em Physics of Plasmas}, 26:032104, 2019.

\bibitem{Morse_PhysFluids_1973}
R.~L. Morse and C.~W. Nielson.
\newblock {Occurrence of high‐energy electrons and surface expansion in
  laser‐heated target plasmas}.
\newblock {\em Physics of Fluids}, 16:909, 1973.

\bibitem{Malone_PhysRevLett_1975}
R.~C. Malone, R.~L. McCrory, and R.~L. Morse.
\newblock {Indications of strongly flux-limited electron thermal conduction in
  laser-target experiments}.
\newblock {\em Physical Review Letters}, 34:721, 1975.

\bibitem{Bell_PhysRevLett_1981}
A.~R. Bell, R.~G. Evans, and D.~J. Nicholas.
\newblock Electron energy transport in steep temperature gradients in
  laser-produced plasmas.
\newblock {\em Physical Review Letters}, 46:243, 1981.

\bibitem{Meezan_PhysPlasmas_2020}
N.~B. Meezan, D.~T. Woods, N.~Izumi, H.~Chen, H.~A. Scott, M.~B. Schneider,
  D.~A. Liedahl, O.~S. Jones, G.~B. Zimmerman, J.~D. Moody, O.~L. Landen, and
  W.~W. Hsing.
\newblock {Evidence of restricted heat transport in National Ignition Facility
  Hohlraums}.
\newblock {\em Physics of Plasmas}, 27:102704, 2020.

\bibitem{Murillo_PhysRevE_2010}
M.~S. Murillo.
\newblock {X-ray Thomson scattering in warm dense matter at low frequencies}.
\newblock {\em Physical Review E}, 81:036403, 2010.

\bibitem{Stanton_PhysRevX_2018}
L.~G. Stanton, J.~N. Glosli, and M.~S. Murillo.
\newblock {Multiscale molecular dynamics for heterogeneous charged systems}.
\newblock {\em Physical Review X}, 8:021044, 2018.

\bibitem{Jian_JComputPhys_2006}
D.~Jian, B.~Fix, J.~Glimm, J.~Xicheng, L.~Xiolin, L.~Yuanhua, and W.~Lingling.
\newblock {A simple package for front tracking}.
\newblock {\em Journal of Computational Physics}, 213(2):613, 2006.

\bibitem{Gittings_ComputSciDisc_2008}
M.~Gittings, R.~Weaver, M.~Clover, T.~Betlach, N.~Byrne, R.~Coker, E.~Dendy,
  R.~Hueckstaedt, K.~New, W.~R. Oakes, D.~Ranta, and R.~Stefan.
\newblock {The RAGE radiation-hydrodynamic code}.
\newblock {\em Computational Science and Discovery}, 1:015005, 2008.

\bibitem{Vold_PhysPlasmas_2017}
E.~L. Vold, R.~M. Rauenzahn, C.~H. Aldrich, K.~Molvig, A.~N. Simikov, and B.~M.
  Haines.
\newblock {Plasma transport in an Eulerian AMR code}.
\newblock {\em Physics of Plasmas}, 24:042702, 2017.

\bibitem{Mihalas_JQSRT_1982}
D.~Mihalas and R.~Weaver.
\newblock {Time-dependent radiative transfer with automatic flux limiting}.
\newblock {\em Journal of Quantitative Spectroscopy and Radiative Transfer},
  28(3):213, 1982.

\bibitem{McGlinchey_thesis_2017}
Kristopher McGlinchey.
\newblock {\em {Radiation-Hydrodynamics Simulations of the Impact of
  Instabilities and Asymmetries on Inertial Confinement Fusion}}.
\newblock PhD thesis, Imperial College London, 2017.

\bibitem{Faik_ComputPhysCommun_2018}
S.~Faik, A.~Tauschwitz, and I.~Iosilevskiy.
\newblock {The equation of state package FEOS for high energy density matter}.
\newblock {\em Computer Physics Communications}, 227:117, 2018.

\bibitem{Atzeni_book}
S.~Atzeni and J.~{Meyer-ter-Vehn}.
\newblock {\em {The Physics of Inertial Fusion}}.
\newblock Oxford University Press, 2004.

\bibitem{Bosch_NuclFusion_1992}
{H.-S. Bosch and G. M. Hale}.
\newblock {Improved fomrulas for fusion cross-sections and thermal
  reactivities}.
\newblock {\em Nuclear Fusion}, 32(4):611, 1992.

\bibitem{Fryxell_AstrophysJSupplSeries_2000}
B.~Fryxell, K.~Olson, P.~Ricker, F.~X. Timmes, M.~Zingale, D.~Q. Lamb,
  P.~MacNeice, R.~Rosner, J.~W. Truran, and H.~Tufo.
\newblock {FLASH: An adaptive mesh hydrodynamics code for modeling
  astrophysical thermonuclear flashes}.
\newblock {\em The Astrophysical Journal: Supplement Series}, 131:273, 2000.

\bibitem{Brown_PhysRevLett_2011}
C.~R.~D. Brown, D.~J. Hoarty, S.~F. James, D.~Swatton, S.~J. Hughes, J.~W.
  Morton, T.~M. Guymer, M.~P. Hill, D.~A. Chapman, J.~E. Andrew, A.~J. Comley,
  R.~Shepherd, J.~Dunn, H.~Chen, M.~Schneider, G.~Brown, P.~Beiersdorfer, and
  J.~Emig.
\newblock {Measurements of electron transport in foils irradiated with a
  picosecond time scale laser pulse}.
\newblock {\em Physical Review Letters}, 106:185003, 2011.

\bibitem{Hoarty_PhysRevLett_2013}
D.~J. Hoarty, P.~Allan, S.~F. James, C.~R.~D. Brown, L.~M.~R. Hobbs, M.~P.
  Hill, J.~W. O. Harris~J. Morton, M.~G. Brookes, R.~Shepherd, J.~Dunn,
  H.~Chen, E.~{Von Marley}, P.~Beiersdorfer, H.~K. Chung, R.~W. Lee, G.~Brown,
  and J.~Emig.
\newblock {Observations of the effect of ionization-potential depression in hot
  dense plasma}.
\newblock {\em Physical Review Letters}, 110:265003, 2013.

\bibitem{Huang_PhysPlasmas_2017}
{C.-K. Huang}, K.~Molvig, B.~J. Albright, E.~S. Dodd, E.~L. Vold, G.~Kagan, and
  N.~M. Hoffman.
\newblock {Study of the ion kinetic effects in ICF runaway burn using a
  quasi-1D hybrid model}.
\newblock {\em Physics of Plasmas}, 24:022704, 2017.

\bibitem{Goldsack_PhysFluids_1982}
T.~J. Goldsack, J.~D. Kilkenny, B.~J. MacGowan, P.~F. Cunningham, C.~L.~S.
  Lewis, M.~H. Key, and P.~T. Rumsby.
\newblock {Evidence of large heat fluxes from the mass ablation rate of
  laser-irradiated spherical targets}.
\newblock {\em Physics of Fluids}, 25:1634, 1982.

\bibitem{Rosen_HEDP_2011}
M.~Rosen, H.~Scott, D.~Hinkel, E.Williams, D.~Callahan, R.~Town, L.~Divol,
  P.~Michel, W.~Kruer, L.~Suter, R.~London, J.~Harte, and G.~Zimmerman.
\newblock {The role of a detailed configuration accounting (DCA) atomic physics
  package in explaining the energy balance in ignition-scale hohlraums}.
\newblock {\em High Energy Density Physics}, 7:180, 2011.

\bibitem{nrl_formulary}
J.~D. Huba.
\newblock {\em {NRL Plasma Formulary}}.
\newblock {The Office of Naval Research}, 2016.

\bibitem{Larsen_JQSRT_1994}
J.~T. Larsen and S.~M. Lane.
\newblock {HYADES - A plasma hydrodynamics code for dense plasmas}.
\newblock {\em Journal of Quantitative Spectroscopy and Radiative Transfer},
  51:179, 1994.

\bibitem{DEIRA_code_manual}
M.~Basko.
\newblock {DEIRA: A 1-D 3-T hydrodynamic code for simulating ICF targets driven
  by fast ion beams}.
\newblock Technical report, Institute for Theoretical and Experimental Physics,
  Moscow.

\bibitem{Ramis_ComputPhysComm_2009}
R.~Ramis, J.~{Meyer-ter-Vehn}, and J.~Ram\'{i}rez.
\newblock {MULTI2D – a computer code for two-dimensional radiation
  hydrodynamics}.
\newblock {\em Computer Physics Communications}, 180:977, 2009.

\bibitem{Pomraning_book}
G.~C. Pomraning.
\newblock {\em {The Equations of Radiation Hydrodynamics}}.
\newblock Courier Corporation, 2005.

\bibitem{Meyer_JComputPhys_2014}
C.~D. Meyer, D.~S. Balsara, and T.~Aslam.
\newblock {A stabilized Runge-Kutta-Legendre method for explicit
  super-time-stepping of parabolic and mixed equations}.
\newblock {\em Journal or Computational Physics}, 257:594, 2014.

\bibitem{Blazek_book}
J.~Blazek.
\newblock {\em {Computational Fluid Dynamics: Principles and Applications}}.
\newblock Butterworth-Heinemann, 2005.

\bibitem{Kadioglu_INL_2008}
S.~Y. Kadioglu, R.~R. Nourgaliev, and V.~A. Mousseau.
\newblock {A comparative study of the harmonic and arithmetic averaging of
  diffusion coefficients for non-linear heat conduction problems}.
\newblock Technical report, Idaho National Laboratory.

\bibitem{Chang_MathComputModelling_1990}
K.~C. Chang and U.~J. Payne.
\newblock {Analytical and numerical approaches for heat conduction in composite
  materials}.
\newblock {\em Mathematical and Computer Modelling}, 14:899, 1990.

\bibitem{Tsui_NumHeatTrans_2014}
{Tsui Y.-Y.}, {Lin S.-W.}, and {Ding K.-J.}
\newblock {Modeling of heat transfer across the interface in two-fluid flows}.
\newblock {\em Numerical Heat Transfer, Part B: Fundamentals}, 66:162, 2014.

\bibitem{Patankar_book}
S.~V. Patankar.
\newblock {\em {Numerical heat transfer and fluid flow}}.
\newblock Hemisphere Publishin Corporation, 1980.

\bibitem{Incropera_book}
F.~P. Incropera and D.~P. {De Witt}.
\newblock {\em {Fundamentals of Heat and Mass Transfer}}.
\newblock Wiley, 2002.

\bibitem{Kamm_LANL_2008}
J.~R. Kamm, J.~S. Brock, S.~T. Brandon, D.~L. Cotrell, B.~Johnson, P.~Knupp,
  W.~J. Rider, T.~G. Truncano, and V.~G. Weirs.
\newblock {Enhanced verifiation testsuite for physics simulation code}.
\newblock Technical Report LA-14379, Los Alamos National Laboratory, 2008.

\bibitem{Zeldovich_book}
Y.~B. Zeldovich and Y.~P. Raizer.
\newblock {\em {Physics of Shock Waves and High Temperature Hydrodynamic
  Phenomena}}.
\newblock Dover, Mineola, NY, 2002.

\bibitem{Reinicke_PhysFluidsA_1991}
P.~Reinicke and J.~{Meyer-ter-Vehn}.
\newblock {The point explosion with heat conduction}.
\newblock {\em Physics of Fluids A: Fluid Dynamics}, 3:1807, 1991.

\bibitem{Shafranov_SovPhysJETP_1957}
V.~D. Shafranov.
\newblock {The structure of shock waves in a plasma}.
\newblock {\em Soviet Physics Journal of Experimental and Theoretical Physics},
  5:1183, 1957.

\bibitem{Maddix_JComputPhys_2018}
D.~C. Maddix, L.~Sampaio, and M.~Gerritsen.
\newblock {Numerical artifacts in the generalized porous medium equation: Why
  harmonic averaging itself is not to blame}.
\newblock {\em Journal of Computational Physics}, 361:280, 2018.

\bibitem{vanderHolst_AstrophysJSupplSeries_2011}
B.~{van der Holst}, G.~T\'{o}th, I.~V. Sokolov, K.~G. Powell, J.~P. Holloway,
  E.~S. Myra, Q.~Stout, M.~L. Adams, J.~E. Morel, S.~Karni, B.~Fryxell, and
  R.~P. Drake.
\newblock {CRASH: A block-adaptive-mesh code for radiative shock hyrdodynamics
  - Implementation and verification}.
\newblock {\em The Astrophysical Journal Supplement Series}, 23:194, 2011.

\bibitem{Spitzer_book}
L.~Spitzer.
\newblock {\em {Physics of Fully Ionized Gases}}.
\newblock Interscience, New York, NY, 1962.

\bibitem{Haack_PhysRevE_2017}
J.~R. Haack, C.~D. Hauck, and M.~S. Murillo.
\newblock {Interfacial mixing in high-energy-density matter with a multiphysics
  kinetic model}.
\newblock {\em Physical Review E}, 96:063310, 2017.

\bibitem{Ross_JChemPhys_1983}
M.~Ross, F.~H. Ree, and D.~A. Young.
\newblock {The equation of state of molecular hydrogen at very high density}.
\newblock {\em The Journal of Chemical Physics}, 79:1487, 1983.

\bibitem{Lee_PhysFluids_1984}
Y.~T. Lee and R.~M. More.
\newblock {An electron conductivity model for dense plasmas}.
\newblock {\em Physics of Fluids}, 27:1273, 1984.

\bibitem{Stanton_PhysRevE_2016}
L.~G. Stanton and M.~S. Murillo.
\newblock {Ionic transport in high-energy-density matter}.
\newblock {\em Physical Review E}, 93:043203, 2016.

\bibitem{Apfelbaum_PhysPlasmas_2020}
E.~M. Apfelbaum.
\newblock {The calculations of thermophysical properties of low-temperature
  gallium plasma}.
\newblock {\em Physics of Plasmas}, 22:092703, 2020.

\bibitem{Hazak_PhysRevE_2001}
G.~Hazak, Z.~Zinamon, Y.~Rosenfeld, and M.~W.~C. {Dharma-wardana}.
\newblock {Temperature relaxation in two-temperature states of dense
  electron-ion systems}.
\newblock {\em Physical Review E}, 64:066411, 2001.

\bibitem{Gericke_JPhysConfSeries_2005}
D.~O. Gericke.
\newblock {Kinetic approach to temperature relaxation in dense plasmas}.
\newblock {\em Journal of Physics: Conference Series}, 11:111, 2005.

\bibitem{Vorberger_HEDP_2014}
J.~Vorberger and D.~O. Gericke.
\newblock {Comparison of electron-ion energy transfer in dense plasmas obtained
  from numerical simulations and quantum kinetic theory}.
\newblock {\em High Energy Density Physics}, 10:1, 2014.

\bibitem{Dharma-wardana_PhysRevLett_1991}
M.~W.~C. {Dharma-wardana}.
\newblock {Nature of coupled-mode contributions to hot electron relaxation in
  semiconductors}.
\newblock {\em Physical Review Letters}, 66:197, 1991.

\bibitem{Dharma-wardana_PhysRevE_1998}
M.~W.~C. {Dharma-wardana} and F.~Perrot.
\newblock {Energy relaxation and the quasiequation of state of a dense
  two-temperature nonequilibrium plasma}.
\newblock {\em Physical Review E}, 3705:58, 1998.

\bibitem{Vorberger_PhysRevE_2010}
J.~Vorberger, D.~O. Gericke, Th. Bornath, and M.~Schlanges.
\newblock {Energy relaxation in dense, strongly coupled two-temperature
  plasma}.
\newblock {\em Physical Review E}, 81:046404, 2010.

\bibitem{Chapman_PhysRevE_2013}
D.~A. Chapman, J.~Vorberger, and D.~O. Gericke.
\newblock {Reduced coupled-mode approach to electron-ion energy relaxation}.
\newblock {\em Physical Review E}, 88:013102, 2013.

\bibitem{Ichimaru_PhysRevA_1985b}
S.~Ichimaru and S.~Tanaka.
\newblock {Theory of interparticle correlations in dense, high-temperature
  plasmas. V. Electric and thermal conductivities}.
\newblock {\em Physical Review A}, 32:1790, 1985.

\bibitem{Kitamura_PhysRevE_1995}
H.~Kitamura and S.~Ichimaru.
\newblock {Electric and thermal resistivities in high-Z plasmas}.
\newblock {\em Physical Review E}, 51:6004, 1995.

\bibitem{Hu_PhysRevE_2014b}
S.~X. Hu, L.~A. Collins, V.~N. Goncharov, T.~R. Boehly, R.~Epstein, R.~L.
  McCrory, and S.~Skupsky.
\newblock {First-principles opacity table of warm dense deuterium for
  inertial-confinement-fusion applications}.
\newblock {\em Physical Review E}, 90:033111, 2014.

\bibitem{Hu_PhysPlasmas_2016}
S.~X. Hu, L.~A. Collins, V.~N. Gocharov, J.~D. Kress, R.~L. McCrory, and
  S.~Skupsky.
\newblock {First-principles investigations on ionization and thermal
  conductivity of polystyrene for inertial confinement fusion applications}.
\newblock {\em Physics of Plasmas}, 23:042704, 2016.

\bibitem{Starrett_HEDP_2020}
C.~E. Starrett, N.~R. Shaffer, D.~Saumon, R.~Perriot, T.~Nelson, L.~A. Collins,
  and C.~Ticknor.
\newblock {Model for the electrical conductivity in dense plasma mixtures}.
\newblock {\em High Energy Density Physics}, 36:100752, 2020.

\bibitem{Shaffer_PhysRevE_2020}
N.~R. Shaffer and C.~A. Starrett.
\newblock {Model of electron transport in dense plasmas spanning temperature
  regimes}.
\newblock {\em Physical Review E}, 101:053204, 2020.

\bibitem{Charakhchyan_JourApplMechTechPhys_1994}
A.~Charakhch'yan.
\newblock {Numerical investigation of deuterium implosion in a conical target}.
\newblock {\em Journal of Applied Mechanics and Technical Physics}, 4:506,
  1994.

\bibitem{Taitano_PhysPlasmas_2018}
W.~T. Taitano, A.~N. Simakov, L.~Chac\'{o}n, and B.~Keenan.
\newblock {Yield degradation in inertial-confinement-fusion implosions due to
  shock-driven kinetic fuel-species stratification and viscous heating}.
\newblock {\em Physics of Plasmas}, 25:056310, 2018.

\bibitem{Molvig_PhysRevLett_2012}
K.~Molvig, N.~M. Hoffman, B.~J. Albright, E.~M. Nelson, and R.~B. Webster.
\newblock {Knudsen Layer Reduction of Fusion Reactivity}.
\newblock {\em Physical Review Letters}, 109:095001, 2012.

\bibitem{Albright_PhysPlasmas_2013}
B.~J. Albright, K.~Molvig, C.-K. Huang, A.~N. Simakov, E.~S. Dodd, N.~M.
  Hoffman, G.~Kagan, and P.~F. Schmit.
\newblock {Revised Knudsen-layer reduction of fusion reactivity}.
\newblock {\em Physics of Plasmas}, 2013.

\bibitem{Kagan_PhysRevLett_2015}
G.~Kagan, D.~Svyatskiy, H.~G. Rinderknecht, M.~J. Rosenberg, A.~B. Zylstra,
  {C.-K.} Huang, and C.~J. McDevitt.
\newblock {Self-similar structure and experimental signatures of suprathermal
  ion distribution in inertial confinement fusion implosions}.
\newblock {\em Physical Review Letters}, 115:105002, 2015.

\bibitem{Ziman_AdvPhys_1967}
J.~M. Ziman.
\newblock {The electron transport properties of pure liquid metals}.
\newblock {\em Advances in Physics}, 16:551, 1967.

\end{thebibliography}
	
	
\end{document}